\documentclass[11pt]{article}

\usepackage[a4paper,margin=1in]{geometry}
\usepackage{amsmath,amssymb,amsthm}
\usepackage{graphicx}
\usepackage{bm}
\usepackage{bbm}
\usepackage{multirow}
\usepackage{float}
\usepackage[dvipsnames]{xcolor}
\usepackage{booktabs}
\usepackage{enumitem}
\usepackage[round,authoryear]{natbib}
\usepackage{authblk}
\usepackage{microtype}
\usepackage{caption}
\usepackage[hidelinks]{hyperref}

\graphicspath{{figures/}}
\newtheorem{proposition}{Proposition}
\newtheorem{definition}{Definition}
\newenvironment{keywords}{\par\medskip\noindent\textbf{Keywords: }}{\par\medskip}
\allowdisplaybreaks

\newcommand{\bv}{{\bf v}}
\newcommand{\bx}{{\bf x}}

\newcommand{\be}{{\bf e}}

\newcommand{\bcm}{{\bf M}}

\newcommand{\bcv}{{\bf V}}
\newcommand{\bcy}{{\bf Y}}
\newcommand{\bcz}{{\bf Z}}
\newcommand{\bcx}{{\bf X}}
\newcommand{\bcw}{{\bf W}}

\newcommand{\bcq}{{\bf Q}}

\newcommand{\bita}{\bm{\eta}}

\newcommand{\balpha}{\bm{\alpha}}

\newcommand{\bfeta}{\bm{\eta}}

\newcommand{\bgamma}{\bm{\gamma}}
\newcommand{\blambda}{\bm{\lambda}}

\newcommand{\bmu}{\bm{\mu}}

\newcommand{\btheta}{\bm{\theta}}

\newcommand{\bpi}{\bm{\pi}}

\newcommand{\bzero}{{\bf 0}}

\newcommand{\logit}{\text{logit}}

\newcommand{\diag}{\text{diag}}

\title{Endogeneity-Aware Cognitive Diagnostic Model for Multidomain Ordinal Assessments}

\author[1]{Zhiyu Huang}
\author[2]{Jing Ouyang\thanks{Corresponding author: \href{mailto:jingoy@hku.hk}{jingoy@hku.hk}}}
\author[1]{Kai Kang\thanks{Corresponding author: \href{mailto:kangk5@mail.sysu.edu.cn}{kangk5@mail.sysu.edu.cn}}}
\affil[1]{Department of Statistics, Sun Yat-sen University, Guangzhou, Guangdong, China}
\affil[2]{Faculty of Business and Economics, University of Hong Kong, Hong Kong, P. R. China}
\date{}

\begin{document}

\maketitle

\begin{abstract}
{Multidomain assessment batteries generate ordinal item responses that are often summarized through latent attribute profiles. Conventional cognitive diagnostic models (CDMs) provide interpretable measurement models for such profiles, but they typically do not represent directed dependence among latent attributes from distinct domains. We propose an endogeneity-aware cognitive diagnostic model (EACDM) for multivariate ordinal assessments. The model combines a block-structured diagnostic measurement component, in which item groups are linked to domain-specific binary attributes through a block-diagonal $Q$-matrix, with a logistic structural component, in which one attribute block is regressed on another block and subject-level covariates while accounting for latent classification uncertainty. This formulation yields a parsimonious framework for studying endogenous relationships among diagnostic attributes without collapsing domain-specific measurement structure. {We establish identifiability conditions for the $Q$-matrix, effective loadings, latent-profile probabilities, and structural coefficients, and develop a Markov chain Monte Carlo algorithm for joint estimation of the measurement and structural components. Simulation studies demonstrate accurate recovery of item parameters, latent structures, and structural coefficients for the proposed EACDM, whereas conventional CDMs can fail to recover the ground truth when endogeneity is present.} We apply the proposed method to Parkinson's disease data to examine how non-motor latent traits relate to motor impairment profiles.}

\end{abstract}


\section{Introduction}
Clinical assessment batteries are extensively used in medicine to measure multiple symptom domains simultaneously. In Parkinson's disease (PD), for example, 
the Movement Disorder Society--Unified Parkinson's Disease Rating Scale (MDS--UPDRS) evaluates both motor and non-motor manifestations~\citep{goetz2008movement}, while Scales for Outcomes in Parkinson's Disease (SCOPA) and Montreal Cognitive Assessment (MoCA) provide complementary information on autonomic and cognitive functioning~\citep{visser2004assessment, nasreddine2005montreal}, respectively. Large cohort studies such as the Parkinson's Progression Markers Initiative (PPMI) routinely collect these item-level ordinal measurements together with neuroimaging and demographic variables, creating an opportunity to study not only overall disease severity but also the domain-specific latent traits and their interrelationship underlying heterogeneous symptom patterns~\citep{marek2011parkinson}. In such settings, reliance on total scores of these assessments can obscure clinically meaningful differences among symptom dimensions and can mask how impairments in one domain relate to impairments in another.

Cognitive diagnostic models (CDMs), also referred to as diagnostic classification models, provide a natural framework for extracting fine-grained latent profiles from multivariate item responses~\citep[e.g.,][]{tatsuoka1983rule,junker2001cognitive,templin2010diagnostic,de2009dina}. 
A central component of a CDM is the $Q$-matrix, which specifies the attributes measured by each assessment item and thereby translates complex response patterns into interpretable profiles of strengths, weaknesses, or symptom dimensions.
Different CDM families have different response mechanisms for how attributes indicated by the $Q$-matrix contribute to item responses.
Classical CDMs include the DINA~\citep{haertel1989using} and DINO models~\citep{templinhenson2006}, which represent conjunctive and disjunctive response mechanisms through two item-specific parameters: slipping and guessing probabilities. Related restricted latent class models, including the NIDA model and the reduced reparameterized unified model~\citep{dibello2012unified}, impose alternative assumptions about how required attributes affect item responses. More flexible frameworks, including the general diagnostic model~\citep{von2019general}, the log-linear cognitive diagnostic model~\citep{henson2009}, and the generalized DINA model~\citep{delatorre2011}, unify many existing CDMs and allow richer main effects and interaction effects among attributes. Although much of the foundational CDM literature was developed for dichotomous item responses, many psychological and clinical assessments use ordered response categories, which has motivated extensions of CDMs to ordinal and polytomous data \citep{von2008general,templin2010diagnostic,culpepper2019exploratory}.  In addition, substantial work has studied $Q$-matrix specification, validation, exploratory estimation, and identifiability, since reliable recovery of latent profiles depends critically on the item--attribute structure \citep{chen2015statistical,xu2017identifiability,xu2018identifying,culpepper2019exploratory}. These developments provide the measurement foundation for using CDMs to recover interpretable latent attribute profiles from complex item response data, which is especially useful in PD research, where patients may share similar total scores while differing substantially in the composition of their motor, autonomic, and cognitive difficulties.

Despite these advances in diagnostic measurement, the primary inferential target in most CDM literature lies in the recovery of latent attribute profiles, item parameters, and item--attribute structures, rather than the structural relationships among the latent attributes themselves. In standard restricted latent class formulations, dependence among attributes can be represented through the joint distribution of latent attribute profiles, but this representation becomes high-dimensional as the number of attributes increases and does not directly yield parsimonious or directed scientific interpretations. Several important extensions have introduced more structured forms of attribute dependence. Higher-order diagnostic models use one or more continuous higher-order latent traits to induce association among binary or polytomous attributes \citep{delatorre2004,culpepperbalamuta2023}. Hierarchical diagnostic classification models and related Bayesian-network approaches encode prerequisite or directed relations among attributes and allow formal evaluation of candidate attribute structures \citep{templinbradshaw2014,hubtemplin2020}. Explanatory CDMs incorporate respondent-level covariates to explain latent attribute mastery or item responses \citep{park2018explanatory,tan2024bayesian}. More recently, \citet{wayman2025} developed a restricted latent class model for polytomous attributes with correlated attribute distributions via multivariate probit formulations and respondent-specific covariates. However, these approaches do not directly address a common clinical setting in which investigators prespecify two substantively distinct measurement domains, recover discrete latent attributes within each domain, and estimate a directed association from one attribute block to the other while accounting for measurement uncertainty.

To address this gap, we develop a novel endogeneity-aware cognitive diagnostic model (EACDM). In particular, motivated by multidomain PD assessment, the model partitions the items into instrument-informed domains and uses a block-diagonal $Q$-matrix to connect each domain to its own set of binary latent attributes. 
One attribute block is modeled through logistic regressions on the other attribute block and subject-level covariates.
The resulting joint formulation integrates a domain-specific diagnostic measurement model with a directed structural model, enabling cross-domain relationships among latent attributes to be estimated while explicitly accounting for their dependence and latent classification uncertainty.
The proposed framework is applicable to ordinal responses and naturally accommodates a mixture of binary and ordinal items.

The central methodological novelty of this work is the joint estimation of domain-specific discrete latent profiles and directed cross-domain relationships while explicitly accounting for latent classification uncertainty. Our contributions are threefold. First, our work extends conventional CDMs by moving beyond latent classification and offering a clinically interpretable framework for characterizing relationships among latent attributes. Second, we establish conditions for the strict and generic identifiability of the \(Q\)-matrix, measurement parameters, latent-profile distribution, and structural parameters, up to blockwise permutations of the attribute labels. Third, under the proposed identifiability conditions, we develop an efficient Markov chain Monte Carlo algorithm for joint posterior inference on the item parameters, latent attributes, $Q$-matrix, and structural coefficients. Furthermore, through extensive simulation studies and an application to PPMI data, the proposed model demonstrates clear advantages over conventional CDMs when latent attributes arise from two distinct domains and exhibit structural dependence.

The remainder of the paper is organized as follows. Section~\ref{sec:model setup}
introduces the EACDM and establishes its identifiability conditions.
Section~\ref{sec:bayesian analysis} presents the Bayesian formulation and MCMC
algorithm. Section~\ref{sec:simu} evaluates the finite-sample performance of the
proposed method and compares it with conventional CDMs. Section~\ref{sec:real data}
applies the method to the PPMI data. Section~\ref{sec:discussion} concludes with
a discussion.

\section{Model Setup}
\label{sec:model setup}

\subsection{Cognitive Diagnostic Measurement Model for Ordinal Responses}

Consider a PD assessment with $J$ items designed to measure $K$ binary latent attributes. For a random subject in the population, denote the observed ordinal responses to $J$ items by $\bcy=(Y_1,\ldots,Y_J)^\top$, where $Y_j\in\{0,\ldots,M_j-1\}$ is the response to item $j$, and $M_j\ge 2$ is the number of ordered response categories. 
The response pattern space is $\mathcal{Y}=\times_{j=1}^J \{0,\ldots,M_j-1\}$.
The subject is associated with a $K$-dimensional binary latent attribute profile
$\balpha=(\alpha_1,\ldots,\alpha_K)^\top\in\{0,1\}^K$, where $\alpha_k=1$ indicates the presence of the $k$th latent attribute and $\alpha_k=0$ indicates its absence. In the PD application, these latent attributes may represent domain-specific impairment statuses related to motor, autonomic, and cognitive functioning.
We index the $2^K$ possible attribute profiles by latent classes: with $\bv=(2^{K-1},2^{K-2},\ldots,1)^\top$, the map $c=\balpha^\top\bv$ assigns each $\balpha\in\{0,1\}^K$ to a unique class label $c\in\{0,\ldots,2^K-1\}$.

The ordinal response is driven by the latent attributes through the following formulation.
For a subject in latent class $c$, we introduce a latent continuous response $Y_j^*$ such that for $m = 0, \ldots, M_j-1$,
\[
Y_j=m
\quad \Longleftrightarrow \quad
\tau_{jc,m}\le Y_j^*<\tau_{jc,m+1},
\]
where $-\infty=\tau_{jc,0}<\tau_{jc,1}<\hdots<\tau_{jc,M_j-1}<\tau_{jc,M_j}=\infty$.
{Following \citet{culpepper2019exploratory}, we fix {$\tau_{jc,m}=m-1$, $m=1,\ldots,M_j-1$,} to avoid weak threshold identification and costly Metropolis--Hastings updates in ordinal mixture models \citep{cowles1996accelerating}.}
{Let $\balpha_c$ denote the binary attribute profile indexed by class $c$, and let $\bcq=(q_{jk})_{J\times K}\in\{0,1\}^{J\times K}$ denote the item--attribute incidence matrix. Its $j$th row is $\boldsymbol q_j=(q_{j1},\ldots,q_{jK})^\top$. {Under the main-effects parameterization, define the effective loading $\delta_{jk}=q_{jk}\beta_{jk}$, where $\beta_{jk}>0$ is estimated only when $q_{jk}=1$ and $\delta_{jk}=0$ when $q_{jk}=0$. Let $\boldsymbol\delta_j=(\delta_{j1},\ldots,\delta_{jK})^\top$, $\boldsymbol\Delta=(\delta_{jk})_{J\times K}$, and $\boldsymbol\beta_0=(\beta_{10},\ldots,\beta_{J0})^\top$.} Conditional on $\balpha^\top\bv=c$, the augmented response is modeled as}
\[
{Y_j^*=\mu_{jc}+\epsilon_j,
\qquad
\mu_{jc}=\beta_{j0}+\sum_{k=1}^Kq_{jk}\beta_{jk}\alpha_{ck}
=\beta_{j0}+\boldsymbol\delta_j^\top\balpha_c,
\qquad
\epsilon_j\sim N(0,1).}
\]
{Here, $\beta_{j0}$ is the item-specific intercept. If $q_{jk}=1$, then $\beta_{jk}>0$ is the active main effect of attribute $k$ on item $j$; if $q_{jk}=0$, then $\delta_{jk}=0$ exactly and an inactive auxiliary $\beta_{jk}$ is not a likelihood parameter. Thus, the measurement object of interest is the intercept vector together with $(\bcq,\boldsymbol\Delta)$, or equivalently $\bcq$ and the active coefficients.}
Therefore, for \(m=0,\ldots,M_j-1\), the class-specific response probability is
\[
\theta_{jcm}
=
P(Y_j=m\mid \balpha^\top\bv=c)
=
{\Phi\!\left(\tau_{jc,m+1}-\mu_{jc}\right)}
-
{\Phi\!\left(\tau_{jc,m}-\mu_{jc}\right),}
\]
where $\Phi(\cdot)$ is the cumulative distribution function of the standard normal distribution. Let $\btheta_{jc}=(\theta_{jc0},\ldots,\theta_{jc,M_j-1})^\top$ denote the conditional response probability vector for item $j$ in class $c$.

\subsection{Structural Model for Endogenous Latent Attributes}
\label{sec:model for attributes}

When one latent attribute is associated with one or more latent attributes, we need to further establish models to account for dependence among the latent attributes $\balpha$.
Without loss of generality, we assume the first $K_1$ latent attributes are endogenous whereas the remaining $K_2$ latent attributes are exogenous and $K_1+K_2=K$.
Denote $\balpha^{(1)}$ and $\balpha^{(2)}$ as endogenous and exogenous latent attributes, respectively, and $\balpha = ((\balpha^{(1)})^{\top},(\balpha^{(2)})^{\top})^\top$.  To assess the interrelationship between latent attributes, we propose logistic models as follows: for $k=1,\hdots,K_1$,
\begin{equation}
\label{eq:logistic}
\operatorname{logit}\!\left\{
\Pr\bigl(\alpha_k^{(1)}=1\mid \boldsymbol\alpha^{(2)}, {\bcz}\bigr)
\right\}
=
\boldsymbol\lambda_k^\top{\bcz}
+
\boldsymbol\gamma_k^\top\boldsymbol\alpha^{(2)},
\end{equation}
where $\logit(p)=\log({p}/({1-p}))$ is the logit link function, 
$\bgamma_k$ is a $K_2$-dimensional vector of unknown parameters capturing how exogenous latent attributes affect the endogenous attribute $\alpha_{k}^{(1)}$, {$\bcz$ is an $R$-dimensional covariate vector, including an intercept, and $\blambda_k$ is the corresponding $R$-dimensional coefficient vector.}
We assume that the components of $\boldsymbol\alpha^{(1)}$ are conditionally independent given $(\boldsymbol\alpha^{(2)}, {\bcz})$, so that Equation~\eqref{eq:logistic}, together with a marginal distribution for $\boldsymbol\alpha^{(2)}$, induces the joint distribution of the complete latent profile $\boldsymbol\alpha$.
We assume a saturated population distribution over profiles for exogenous latent attributes. For each $\mathbf a\in\{0,1\}^{K_2}$,
let
$\pi_{\mathbf{a}}^{(2)}=\operatorname{Pr}(\boldsymbol{\alpha}^{(2)}=\mathbf{a})$ be 
the population prevalence of exogenous latent profile $\mathbf a$.
We collect these probabilities in {$\bpi^{(2)}=(\pi_{\mathbf a}^{(2)}:\mathbf a\in\{0,1\}^{K_2})^{\top}$.}
For notational convenience, we combine the covariate coefficients and the exogenous-attribute coefficients into $\boldsymbol{\eta}_k
=
\big(\blambda_k^\top,\bgamma_k^\top\big)^\top$. {These structural coefficients are collected as $\blambda=(\blambda_1,\ldots,\blambda_{K_1})$, $\bgamma=(\bgamma_1,\ldots,\bgamma_{K_1})$ and $\bita = (\bita_1, \ldots, \bita_{K_1})$.}

{To encode the nonoverlapping domain structure, we partition the previously defined matrices $\bcq$ and $\boldsymbol\Delta$ conformably as}
\begin{align*}
{\bcq=   \begin{pmatrix}
\bcq_1 & \bzero \\
\bzero& \bcq_2
\end{pmatrix},\qquad
\boldsymbol\Delta=\begin{pmatrix}
\boldsymbol\Delta_1 & \bzero \\
\bzero & \boldsymbol\Delta_2
\end{pmatrix}.}
\end{align*}
{Here, $\bcq_1$ and $\bcq_2$ are $J_1\times K_1$ and $J_2\times K_2$ matrices, respectively. The first $J_1$ items therefore measure only the $K_1$ endogenous attributes, whereas the remaining $J_2$ items measure only the $K_2$ exogenous attributes.}

This structural specification is motivated by settings where latent traits naturally fall into two related domains. In the PPMI study, for example, latent motor symptoms may depend on latent non-motor symptoms. Under this interpretation, items from the MDS-UPDRS primarily measure the motor domain, whereas items from the SCOPA-AUT and MoCA primarily measure non-motor domains. The block-diagonal $Q$-matrix encodes this domain-specific measurement structure, while the logistic model in \eqref{eq:logistic} captures directed dependence from the exogenous latent attributes to the endogenous ones.

\subsection{Identifiability Conditions}

Because identifiability is required for valid estimation and inference, we
first formalize its strict and generic forms for the EACDM.

\begin{definition}

The EACDM parameter 
$(\boldsymbol\beta_0,\bcq,\boldsymbol\Delta,\bm\pi^{(2)},\bgamma,\blambda)$
is strictly identifiable if any two sets of parameter values induce the same
observed-response distribution for all covariate values if and only if they are
identical up to simultaneous permutations of the latent-attribute labels within
each block\footnote{Such permutations correspond to the same permutations of the
columns of $\bcq_d$ and $\boldsymbol\Delta_d$ and the associated entries of the
structural coefficients.}. Generic identifiability means that this property
holds except for a set of parameter values of Lebesgue
measure zero.

\end{definition}

\noindent

Model identifiability is nontrivial in the proposed EACDM because both the latent attributes and their structural dependence are unobserved. {Without additional constraints, distinct measurement structures, effective loadings, exogenous-profile probabilities, and structural coefficients may induce the same distribution of the observed ordinal responses.} We therefore impose conditions on both the measurement and structural components.
\begin{enumerate}[label=($A${\arabic*}),leftmargin=1.5cm]
     \item \label{A1} $\prod_{j=1}^{J} M_j-1 \geq 2^K(\sum_{j=1}^{J} M_j -J) + 2^K - 1 $;
    \item \label{A2} {The parameters $\bgamma$ and $\blambda$ and the covariate vectors $\bcz_i$, $i=1,\ldots,N$, are finite;}
   
    \item \label{A3} {The covariate design matrix with rows $\bm{Z}_{1}^{T},\ldots,\bm{Z}_{N}^{T}$ has full column rank;}

\item \label{A4} {The exogenous latent-profile distribution has full support: $\pi_{\mathbf a}^{(2)}>0$ for every $\mathbf a\in\{0,1\}^{K_2}$. Equivalently, $\mathcal S=\{0,1\}^{K_2}$; in particular, $\mathcal S$ contains $\mathbf0$ and the basis vectors $\be_1,\ldots,\be_{K_2}$.}
\end{enumerate}

{Condition $(A1)$ is a necessary dimensional check.}
{Conditions $(A2)$ and $(A3)$ are regularity requirements for the structural regression. Condition $(A4)$, together with the finite logistic coefficients in $(A2)$, gives positive probability to every complete latent profile and also supplies the zero and basis exogenous profiles needed to separate the components of $\bgamma$. } We now present sufficient conditions for strict and generic identifiability results for the proposed EACDM.

\begin{proposition} [Strict Identifiability]
 \label{covariate strict identifiability CDMs}
{Suppose $(A1)$--$(A4)$ hold. If each block $\bcq_d$, $d=1,2$, satisfies Condition $(A5)$ below, then $(\boldsymbol\beta_0,\bcq,\boldsymbol\Delta,\bm\pi^{(2)},\bgamma,\blambda)$ is strictly identifiable. }
 	 	\begin{enumerate}[label=(A\arabic*),leftmargin=1.5cm]
 	 	 	 	 \setcounter{enumi}{4}
\item \label{C4star} {After a row permutation, each block has the form $\bcq_d=(\mathbf I_{K_d}^\top,\mathbf I_{K_d}^\top,(\bcq_d^*)^\top)^\top$, containing two $K_d\times K_d$ identity matrices and one submatrix $\bcq_d^*$. Moreover, for every pair of distinct within-block profiles, at least one item represented in $\bcq_d^*$ has different category-response probability vectors under those profiles.}
 		\end{enumerate}

 \end{proposition}

 \begin{proposition} [Generic Identifiability]
 \label{generic CDMs}
{Suppose $(A1)$--$(A4)$ hold. If each block $\bcq_d$, $d=1,2$, satisfies Condition $(A5')$ below, then $(\boldsymbol\beta_0,\bcq,\boldsymbol\Delta,\bm\pi^{(2)},\bgamma,\blambda)$ is generically identifiable.}

 		\begin{enumerate}[label=($A$\arabic*$^{\prime}$),leftmargin=1.5cm]
 		 	\setcounter{enumi}{4}
\item \label{C4prime2} {After a row permutation, each block can be written as $\bcq_d=((\bcq_d^{[1]})^\top,(\bcq_d^{[2]})^\top,(\bcq_d^*)^\top)^\top$, containing one submatrix $\bcq_{(J_d-2K_d) \times K_d}^{*}$ in which each attribute is required by at least one item  and, for $r=1,2$,}
 			\begin{equation}
 			\label{Q_i}
{\bcq_d^{[r]} = \left(\begin{array}{cccc}
						1& * & \cdots & * \\  
						*& 1 &  \cdots &* \\
						\vdots & \vdots & \ddots & \vdots \\
						* & * & \cdots & 1 \\
 					\end{array} \right),\quad r =1,2,\quad d=1,2,}
 			\end{equation}
	 		{where \(*\) denotes either zero or one.}
		\end{enumerate}
 \end{proposition}

{Under the $Q$-restricted likelihood, Conditions~\ref{C4star} and~\ref{C4prime2} are imposed separately on the two measurement blocks and are adapted from restricted-latent-class identifiability results \citep{xu2018identifying,gu2018sufficient,culpepper2019exploratory}. Once the class-response probabilities are identified with their blockwise attribute labels, the fixed thresholds and injective probit link identify every class-specific linear predictor. The all-zero profile identifies the item intercept $\beta_{j0}$, and the
contrast between the all-zero profile and the profile with only attribute $k$
present identifies the effective loading $\delta_{jk}$. Positivity gives $q_{jk}=I(\delta_{jk}>0)$ and $\beta_{jk}=\delta_{jk}$ for active entries. The identified joint class probabilities then determine the structural conditional probabilities, and Conditions $(A2)$--$(A4)$ identify $(\bm\pi^{(2)},\bgamma,\blambda)$. Full proofs are provided in the Supplementary Material.}

\section{Bayesian Analysis}
\label{sec:bayesian analysis}

This section presents the Bayesian formulation and posterior computation for
the EACDM. For independent subjects $i=1,\ldots,N$, let
$\bcy_i=(Y_{i1},\ldots,Y_{iJ})^\top$ and
$\bcy_i^*=(Y_{i1}^*,\ldots,Y_{iJ}^*)^\top$ denote the observed ordinal and
augmented continuous response vectors, respectively. Each subject has a latent
attribute profile
$\balpha_i=((\balpha_i^{(1)})^\top,(\balpha_i^{(2)})^\top)^\top$
and corresponding class label $c_i=\balpha_i^\top\bv$. We first specify the
prior distributions and then describe the MCMC algorithm.

\subsection{Prior Specification}

\begin{figure}[!htbp]
\centering
\includegraphics[width=\linewidth,height=0.72\textheight,keepaspectratio]{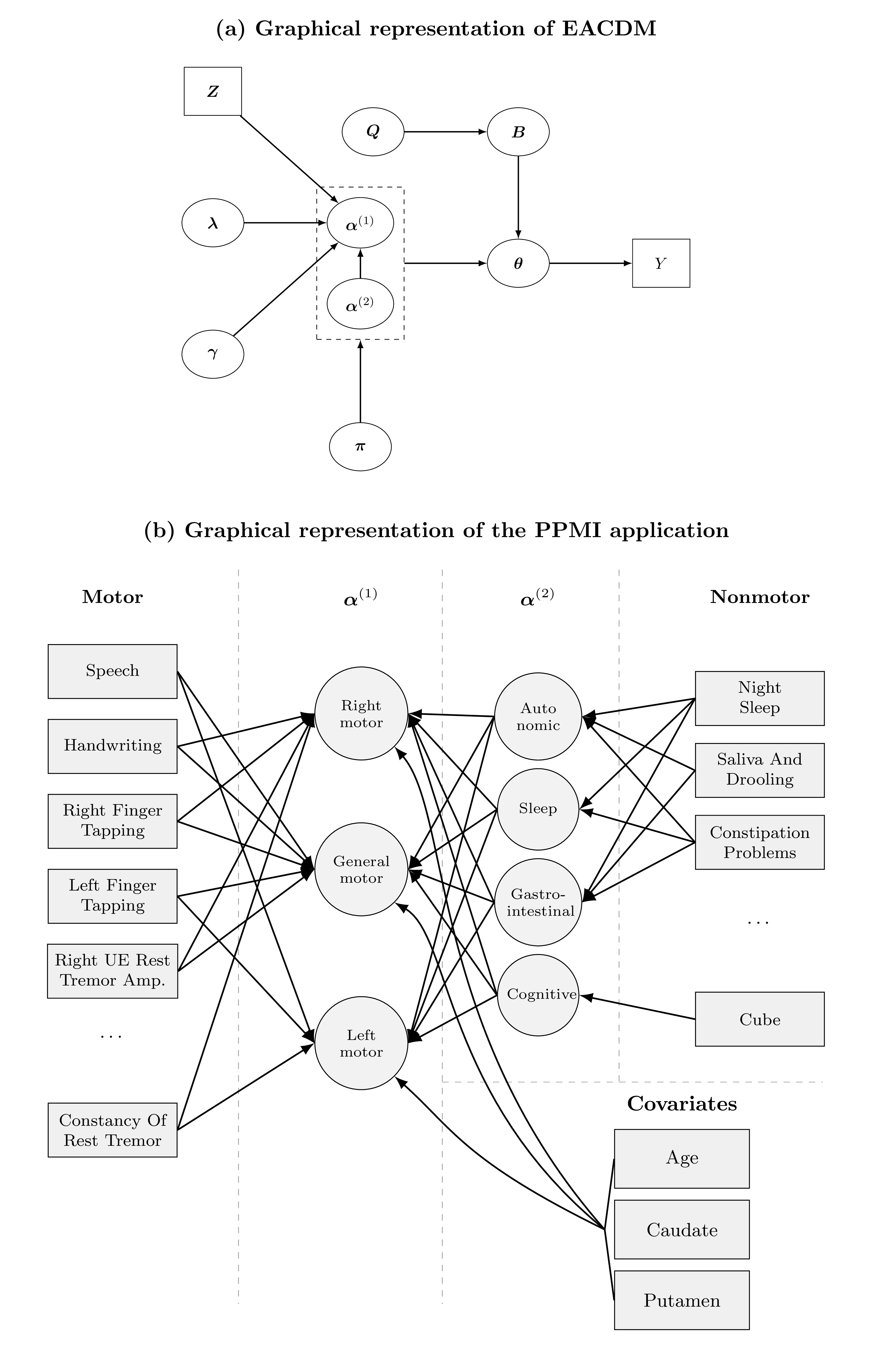}
\caption{Panel (a) shows the path diagram of the proposed EACDM. Solid arrows indicate stochastic dependencies, and the dashed box groups the two latent attribute blocks, \(\alpha^{(1)}\) and \(\alpha^{(2)}\). Panel (b) illustrates the estimated item--attribute and structural relationships in the PPMI data application. Rectangular nodes denote observed variables, including motor items, non-motor items, and covariates; circular nodes denote latent attributes. Directed arrows from latent attributes to items represent nonzero entries in the estimated \(Q\)-matrices. Directed arrows from \(\alpha^{(2)}\) and the covariates to \(\alpha^{(1)}\) represent the structural component of the model.}
\label{fig:para_graph}
\end{figure}

Figure~\ref{fig:para_graph}(a) summarizes the Bayesian path diagram for the EACDM, and Figure~\ref{fig:para_graph}(b) illustrates the estimated item--attribute and structural relationships in the PPMI application.

For the structural model, we assign independent multivariate normal priors $\bita_k\sim N(\bmu_{\eta},\bcm_{\eta})$, for $k = 1, ..., K_1$. Recall that $\bpi^{(2)}
=
(\pi_{\mathbf a}^{(2)}:
\mathbf a\in\{0,1\}^{K_2})^{\top}$ contains the population probabilities of the exogenous latent profiles.
We assign the conjugate prior $\bpi^{(2)}\sim\text{Dirichlet}({\bf n}_0)$,
 with ${\bf n}_0$ set to a vector of ones in the analyses.
 The latent-attribute prior contribution used in the sampler can be written as
\[
{p(\balpha_i\mid\bpi^{(2)},\bita,\bcz_i)
=
\pi_{\balpha_i^{(2)}}^{(2)}
p(\balpha_i^{(1)}\mid\balpha_i^{(2)},\bcz_i,\bita).}
\]
For notational convenience, let
$\mathcal J_1=\{1,\ldots,J_1\}$ and
$\mathcal J_2=\{J_1+1,\ldots,J\}$ denote the item sets for the two measurement
blocks. For $d=1,2$ and $j\in\mathcal J_d$, write $\beta_{j0}^{(d)}$ and
$\delta_{jk}^{(d)}$ for the block-specific item intercept and effective loading
introduced in Section~\ref{sec:model setup}.
{We assign the unconstrained intercept the prior}
\(
{\beta_{j0}^{(d)}\sim N(0,\sigma_{0d}^2).}
\)
{For each main effect $k=1,\ldots,K_d$, we instead use an exact spike-and-positive-slab prior for the effective loading,}
\[
{\delta_{jk}^{(d)}=q_{jk}^{(d)}\beta_{jk}^{(d)},\qquad
\delta_{jk}^{(d)}\mid q_{jk}^{(d)}
\sim
(1-q_{jk}^{(d)})\,\delta_0
+q_{jk}^{(d)}\,N_{(0,\infty)}(0,\sigma_{\beta d}^2),}
\]
{where $\delta_0$ is a point mass at zero and $N_{(0,\infty)}$ denotes a normal distribution truncated to the positive half-line. Equivalently, $\beta_{jk}^{(d)}\sim N_{(0,\infty)}(0,\sigma_{\beta d}^2)$ only when $q_{jk}^{(d)}=1$; no inactive auxiliary coefficient is estimated when $q_{jk}^{(d)}=0$. The inclusion indicators and block-specific inclusion probabilities satisfy}
\[
{q_{jk}^{(d)}\mid\omega_d\sim\operatorname{Bernoulli}(\omega_d),
\qquad
\omega_d\sim\operatorname{Beta}(a_\omega,b_\omega),
\qquad d=1,2.}
\]

\subsection{Posterior Inference}
The Bayesian formulation described in the previous section allows for the implementation of a Gibbs sampler procedure for iteratively updating model parameters. Up to proportionality, the joint posterior is
\begin{equation}
\begin{aligned}
{p(\bcy^*,\balpha,\bpi^{(2)},\boldsymbol\beta_0,\boldsymbol\Delta,\bcq,\bm{\omega},\bita\mid \bcy)}
&\propto
p(\bcy\mid \bcy^*)
{p(\bcy^*\mid \balpha,\boldsymbol\Delta,\boldsymbol\beta_0)}
p(\balpha^{(1)}\mid \balpha^{(2)},\bcz,\bita)\\
&\quad \times
p(\balpha^{(2)}\mid \bpi^{(2)})
p(\bpi^{(2)})
{p(\boldsymbol\Delta,\boldsymbol\beta_0\mid \bcq)}
p(\bcq\mid \bm\omega)
p(\bm\omega)
p(\bita).
\end{aligned}
\label{eq:joint_posterior}
\end{equation}
{We approximate this posterior using a partially collapsed Gibbs sampler with data augmentation. The required updates are stated below.}

\medskip
\noindent\textbf{Step 1. Update the latent attributes and class probabilities.}
For each subject, the latent attribute vector is sampled from a discrete distribution over all $2^K$ possible profiles. Let $\balpha_c=((\balpha_c^{(1)})^\top,(\balpha_c^{(2)})^\top)^\top$ 
such that $\balpha_c^\top\bv=c$. We define
$
p_{ik}(c)
=
\logit^{-1}
\left\{
\bita_k^\top
(\bcz_i^\top,(\balpha_c^{(2)})^\top)^\top
\right\}, $ $ k=1,\ldots,K_1,
$
{and}
\(
\varpi_{ic}
=
\pi_{\boldsymbol\alpha_c^{(2)}}^{(2)}
\prod_{k=1}^{K_1}
p_{ik}(c)^{\alpha_{ck}^{(1)}}
\{1-p_{ik}(c)\}^{1-\alpha_{ck}^{(1)}}.
\)
Here, $\varpi_{ic}$ is the conditional prior probability that subject $i$
belongs to complete latent class $c$. The full conditional distribution of
$c_i$ is
\begin{align}
{P(\balpha_i^{\top}\bv=c\mid \bcy_i,\bpi^{(2)},\boldsymbol\Delta,\bita)}
=
\frac{
\varpi_{ic}\prod_{j=1}^J\theta_{jc,Y_{ij}}
}{
\displaystyle
\sum_{c'=0}^{2^K-1}
\varpi_{ic'}\prod_{j=1}^J\theta_{jc',Y_{ij}}
}.\label{eq:alpha_full_conditional}
\end{align}
{Because $\bpi^{(2)}$ contains only the $2^{K_2}$ exogenous-profile probabilities, define $n_{\mathbf a}^{(2)}=\sum_{i=1}^{N}I(\balpha_i^{(2)}=\mathbf a)$ for $\mathbf a\in\{0,1\}^{K_2}$. Its conjugate update is
\[
\bpi^{(2)}\mid\balpha_1^{(2)},\ldots,\balpha_N^{(2)}
\sim\operatorname{Dirichlet}\!\left({\mathbf n_0}+\{n_{\mathbf a}^{(2)}:\mathbf a\in\{0,1\}^{K_2}\}\right).
\]
}

\medskip
\noindent\textbf{Step 2. Update the augmented responses.}
Let $c_i=\balpha_i^\top\bv$. The augmented responses are updated from truncated
normal distributions,
\begin{equation*}
{Y_{ij}^*\mid Y_{ij}=m,\balpha_i,\bcq,\boldsymbol\Delta,\boldsymbol\beta_0}
\sim
{N\!\left(\beta_{j0}+\sum_{k=1}^K\delta_{jk}\alpha_{ik},\; 1\right)}
I(\tau_{jc_i,m}\le Y_{ij}^*< \tau_{jc_i,m+1}).
\label{eq:ystar_full_conditional}
\end{equation*}

\medskip
\noindent\textbf{Step 3. Update the structural coefficients.}
Let $\bx_i=(\bcz_i^\top,(\balpha_i^{(2)})^\top)^\top $ and let $\mathbf X$ be the $N\times(R+K_2)$ matrix with rows
$\mathbf x_i^\top$. Motivated by Pólya--Gamma representation in
\cite{polson2013bayesian}, we introduce independent
Pólya--Gamma variables $\rho_{ik}\mid\bita_k\sim\mathcal{PG}(1,\bx_i^\top\bita_k)$ for $k = 1, ..., K_1$, and let $\boldsymbol{\rho}_k=(\rho_{1k},\ldots,\rho_{Nk})^\top$, $\bcw_k=\diag(\rho_{1k},\ldots,\rho_{Nk})$
and
$
\boldsymbol \kappa_k=(\alpha_{1k}^{(1)}-1/2,\ldots,\alpha_{Nk}^{(1)}-1/2)^\top.$
Then
\(
\bita_k\mid\balpha^{(1)},\balpha^{(2)},\bcz,\boldsymbol \rho_k
\sim
N(\bmu_{\eta k},\bcv_{\eta k}),
\)
where
\(
    \bcv_{\eta k}
=
(\bcx^\top\bcw_k\bcx+\bcm_\eta^{-1})^{-1}\), 
\(
\bmu_{\eta k}
=
\bcv_{\eta k}(\bcx^\top\boldsymbol \kappa_k+\bcm_\eta^{-1}\bmu_\eta).
\)

\medskip
\noindent\textbf{Step 4. Update the measurement parameters and $Q$-matrices.}
{Given the current latent profiles and augmented responses, update the measurement component block by block. For $d=1,2$, $j\in\mathcal J_d$, and $k=1,\ldots,K_d$, define the partial residual $r_{ijk}^{(d)}=Y_{ij}^*-\beta_{j0}^{(d)}-\sum_{h\ne k}q_{jh}^{(d)}\beta_{jh}^{(d)}\alpha_{ih}^{(d)}$. Conditional on all remaining quantities, let}
\[
{m_{jk}^{(d)}(0)=\prod_{i=1}^N\phi(r_{ijk}^{(d)}),\qquad
m_{jk}^{(d)}(1)=\int_0^\infty
\left\{\prod_{i=1}^N\phi(r_{ijk}^{(d)}-b\alpha_{ik}^{(d)})\right\}
f_+(b;0,\sigma_{\beta d}^2)\,db,}
\]
{where $\phi$ is the standard normal density and $f_+$ is the positive-normal slab density. The collapsed inclusion update is}
\[
{P(q_{jk}^{(d)}=1\mid\text{rest})=
\frac{\omega_d m_{jk}^{(d)}(1)}
{\omega_d m_{jk}^{(d)}(1)+(1-\omega_d)m_{jk}^{(d)}(0)}.}
\]
{Immediately after sampling $q_{jk}^{(d)}$, set $\delta_{jk}^{(d)}=0$ if $q_{jk}^{(d)}=0$. If $q_{jk}^{(d)}=1$, sample its active magnitude before proceeding to the next coordinate:}
\[
{\beta_{jk}^{(d)}\mid q_{jk}^{(d)}=1,\text{rest}
\sim N_{(0,\infty)}\!\left(\widetilde m_{jk}^{(d)},\widetilde v_{jk}^{(d)}\right),\;
\widetilde v_{jk}^{(d)}=\big\{\sum_{i=1}^N(\alpha_{ik}^{(d)})^2+\sigma_{\beta d}^{-2}\big\}^{-1},\;
\widetilde m_{jk}^{(d)}=\widetilde v_{jk}^{(d)}\sum_{i=1}^N\alpha_{ik}^{(d)}r_{ijk}^{(d)},}
\]
{and set $\delta_{jk}^{(d)}=\beta_{jk}^{(d)}$. After cycling through the loading coordinates, update the intercept from}
\[
{\beta_{j0}^{(d)}\mid\text{rest}\sim
N\!\left(\widetilde m_{j0}^{(d)},\widetilde v_{j0}^{(d)}\right),\quad
\widetilde v_{j0}^{(d)}=(N+\sigma_{0d}^{-2})^{-1},\quad
\widetilde m_{j0}^{(d)}=\widetilde v_{j0}^{(d)}\sum_{i=1}^N\left\{Y_{ij}^*-\sum_{k=1}^{K_d}\delta_{jk}^{(d)}\alpha_{ik}^{(d)}\right\}.}
\]
{This ordering defines a valid coordinate-wise partially collapsed Gibbs sweep. The block-specific inclusion probability retains the conjugate update}
\[
\omega_d\mid\bcq_d
\sim
\mathrm{Beta}
\left(
a_\omega+\sum_{j\in\mathcal{J}_d}\sum_{k=1}^{K_d}q_{jk}^{(d)},
\ 
b_\omega+J_dK_d-\sum_{j\in\mathcal{J}_d}\sum_{k=1}^{K_d}q_{jk}^{(d)}
\right),
\qquad d=1,2.
\]

\subsection{Model Selection}
{We compared candidate values of $(K_1,K_2)$ using the following posterior-averaged complete-data information criterion, which was referred to as a modified BIC in the numerical studies:}
{
\begin{equation}
\operatorname{BIC}_{\mathrm{mod}}
=p_{K_1,K_2}\log N
-
\frac{2}{T}\sum_{t=1}^{T}\sum_{i=1}^{N}
\log L_{i,\mathrm{comp}}^{(t)},
\label{eq:modified_bic}
\end{equation}
where
\begin{align}
L_{i,\mathrm{comp}}^{(t)}
={}&
\pi_{\balpha_i^{(2)(t)}}^{(2)(t)}
\prod_{j=1}^{J}
\theta_{j c_i^{(t)},Y_{ij}}^{(t)}
\prod_{k=1}^{K_1}
\{p_{ik}^{(t)}\}^{\alpha_{ik}^{(1)(t)}}
\{1-p_{ik}^{(t)}\}^{1-\alpha_{ik}^{(1)(t)}}.
\label{eq:modified_bic_complete}
\end{align}
Equation~\eqref{eq:modified_bic} averages the complete-data log likelihood over retained posterior draws; it is not the standard observed-data BIC evaluated at a point estimate.}

\section{Simulation}\label{sec:simu}

{We conducted simulation studies to evaluate the finite-sample performance of the proposed EACDM. The measurement targets are the two block-specific $Q$-matrices and the effective loading matrices $\boldsymbol\Delta_d$, $d=1,2$, the structural coefficient matrix $\bfeta$, and the selected latent dimensions $(K_1,K_2)$. We also compare the proposed EACDM with a conventional CDM under strong cross-block dependence among the latent attributes.}

\subsection{Simulation Setting}

Ordinal responses were generated from the full EACDM described in
Section~\ref{sec:model setup}. For subject $i$, the latent attribute vector was 
$\balpha_i=\big((\balpha_i^{(1)})^\top,(\balpha_i^{(2)})^\top\big)^\top$, where $\balpha_i^{(1)}=(\alpha_{i1}^{(1)},\ldots,\alpha_{iK_1}^{(1)})^\top$ denotes the endogenous block and $\balpha_i^{(2)}=(\alpha_{i1}^{(2)},\ldots,\alpha_{iK_2}^{(2)})^\top$ denotes the exogenous block. We considered balanced settings with $K_1=K_2=K$ and  $K=2,3,4$. All items had three ordered response categories, so $M_j=3$ for every item.

The true $Q$-matrix was block diagonal with blocks $\mathbf Q_1$ and $\mathbf Q_2$,
where $\mathbf Q_1$ is a $J_1\times K_1$ matrix for the endogenous item block and
$\mathbf Q_2$ is a $J_2\times K_2$ matrix for the exogenous item block. We considered
$J_1=J_2=24$ and $J_1=J_2=36$, corresponding to total item lengths $J=48$ and
$J=72$, respectively. For each $K$, the same construction was used for both blocks.
Specifically, let $\mathbf C_K$ denote the matrix whose rows contain all two-attribute
combinations among the $K$ attributes; equivalently, each row of $\mathbf C_K$ has
exactly two entries equal to one. The base pattern was
{$\widetilde{\bcq}_K=(\mathbf I_K^\top,\mathbf I_K^\top,\mathbf I_K^\top,\mathbf{C}_K^\top)^\top$}.
Thus, each attribute was measured by at least three single-attribute items, and the design also contained items involving pairs of attributes. The matrices $\bcq_1$ and $\bcq_2$ were obtained by repeating the rows of $\widetilde{\bcq}_K$ cyclically until $J_d$ rows were obtained. For example, when $K=3$ and $J_1=J_2=36$, we have
\[
\mathbf{C}_3=
\begin{pmatrix}
1 & 1 & 0\\
1 & 0 & 1\\
0 & 1 & 1
\end{pmatrix},
\]
 and the 12-row base pattern {$\widetilde{\bcq}_3=(\mathbf I_3,\mathbf I_3,\mathbf I_3,\mathbf{C}_3)^\top$} was repeated three times for $\bcq_1$ and $\bcq_2$. {For the measurement parameters, all item intercepts were set to $-0.2$ and all active magnitudes were set to one. Consequently, the true effective loading matrices were $\boldsymbol\Delta_1=\bcq_1$ and $\boldsymbol\Delta_2=\bcq_2$. Ordinal responses were generated according to $Y_{ij}^*=\beta_{j0}^{(d)}+\sum_k q_{jk}^{(d)}\beta_{jk}^{(d)}\alpha_{ik}^{(d)}+\epsilon_{ij}$.}

The exogenous block $\balpha_i^{(2)}$ was sampled uniformly from $\{0,1\}^K$. We generated an independent binary covariate $Z_i\sim\mathrm{Bernoulli}(0.5)$ for each subject. Conditional on $\balpha_i^{(2)}$ and $Z_i$, the endogenous attributes in $\balpha_i^{(1)}$ were generated independently according to the logistic structural model in \eqref{eq:logistic}. Thus, the second attribute block $\balpha_i^{(2)}$ and the covariate $Z_i$ jointly determined the probabilities of the first attribute block $\balpha_i^{(1)}$. The true values of $\bfeta$ under different $K$ were provided in the Supplementary Material. 
We note that the true parameters in all simulation settings satisfy the conditions of Proposition 1, thereby guaranteeing strict identifiability of the EACDM.

We considered sample sizes $n\in \{500,1000,2000\}$. For each combination of $n$, $J$, and $K$, 100 independent replicates were generated. For each replicate, the MCMC chain was run for 3000 iterations, with the first 2000 iterations discarded as burn-in. The convergence diagnostics for the MCMC algorithm are provided in the Supplementary Material. {For the binary $Q$-matrix, we use the Adjusted Rand Index (ARI; \citealp{rand1971objective}) after aligning attribute labels within each block. For continuous measurement parameters, the appropriate target is $\operatorname{RMSE}(\boldsymbol\Delta)$, computed from posterior estimates of $\boldsymbol\Delta=\bcq\odot\mathbf B_A$ and its true value after the same label alignment; RMSE of inactive auxiliary coefficients is undefined as a likelihood-recovery target. We separately compute RMSE for $\boldsymbol\eta$. For each scenario, we report the median and interquartile range (IQR) of these metrics over 100 replicates.}

\subsection{Results}
{Table~\ref{tab:simulation_results_extended} summarizes the parameter-recovery results. Recovery of the item-attribute latent structure improved with sample size. Across the six settings with $n=2000$, the median ARI ranged from 0.947 to 0.967, compared with 0.825 to 0.879 when $n=500$. The corresponding IQR also generally narrowed as $n$ increased, indicating more stable recovery across replicates. Increasing the number of items generally improved recovery,
although the improvement was not uniform across all settings.

The effective loadings were accurately estimated. For every $(J,K)$
combination, the median RMSE of $\boldsymbol{\Delta}$ decreased monotonically as $n$ increased. For example, when $K=3$ and $J=48$, the median RMSE decreased from 0.090 at $n=500$ to 0.042 at $n=2000$; when $K=3$ and $J=72$, it decreased from 0.081 to 0.039. The corresponding IQRs also decreased, indicating increasingly accurate and stable estimation of the effective loadings.

The structural coefficients had larger RMSE values than the effective loadings, particularly as $K$ increased. This pattern likely reflects the
growth in both the number of possible latent profiles and the number of structural coefficients. Nevertheless, the median RMSE of $\boldsymbol{\eta}$ decreased monotonically with $n$ in every $(J,K)$ setting. For instance, when $K=4$ and $J=72$, the median RMSE decreased from 0.263 at $n=500$ to 0.140 at $n=2000$. Overall, the proposed Bayesian estimation procedure recovered the directed dependence between the two latent attribute blocks, with accuracy improving as the available information increased.}

\begin{table}[ht]
\centering
\caption{Simulation results for different sample sizes, item lengths, and latent dimensions}
\resizebox{\textwidth}{!}{
\begin{tabular}{cccccccccc}
\toprule
& & & & \multicolumn{2}{c}{$ARI(\mathbf{Q})$}
& \multicolumn{2}{c}{$RMSE(\boldsymbol{\Delta})$}
& \multicolumn{2}{c}{$RMSE(\boldsymbol{\eta})$} \\
\cmidrule(lr){5-6}
\cmidrule(lr){7-8}
\cmidrule(lr){9-10}
$n$ & $J$ & $K_1$ & $K_2$
& median & IQR
& median & IQR
& median & IQR \\
\midrule
500  & 48 & 2 & 2 & 0.825 & 0.121 & 0.099 & 0.014 & 0.233 & 0.087 \\
500  & 72 & 2 & 2 & 0.843 & 0.104 & 0.093 & 0.010 & 0.200 & 0.074 \\
1000 & 48 & 2 & 2 & 0.889 & 0.082 & 0.068 & 0.008 & 0.165 & 0.052 \\
1000 & 72 & 2 & 2 & 0.920 & 0.072 & 0.065 & 0.006 & 0.143 & 0.046 \\
2000 & 48 & 2 & 2 & 0.961 & 0.069 & 0.048 & 0.007 & 0.119 & 0.042 \\
2000 & 72 & 2 & 2 & 0.947 & 0.054 & 0.045 & 0.005 & 0.097 & 0.040 \\
\midrule
500  & 48 & 3 & 3 & 0.839 & 0.127 & 0.090 & 0.012 & 0.262 & 0.068 \\
500  & 72 & 3 & 3 & 0.862 & 0.096 & 0.081 & 0.008 & 0.221 & 0.056 \\
1000 & 48 & 3 & 3 & 0.901 & 0.088 & 0.062 & 0.008 & 0.178 & 0.048 \\
1000 & 72 & 3 & 3 & 0.923 & 0.059 & 0.057 & 0.005 & 0.163 & 0.050 \\
2000 & 48 & 3 & 3 & 0.960 & 0.049 & 0.042 & 0.006 & 0.139 & 0.045 \\
2000 & 72 & 3 & 3 & 0.967 & 0.051 & 0.039 & 0.004 & 0.120 & 0.035 \\
\midrule
500  & 48 & 4 & 4 & 0.869 & 0.103 & 0.087 & 0.010 & 0.313 & 0.062 \\
500  & 72 & 4 & 4 & 0.879 & 0.065 & 0.080 & 0.008 & 0.263 & 0.064 \\
1000 & 48 & 4 & 4 & 0.923 & 0.072 & 0.058 & 0.008 & 0.223 & 0.049 \\
1000 & 72 & 4 & 4 & 0.921 & 0.088 & 0.055 & 0.006 & 0.196 & 0.043 \\
2000 & 48 & 4 & 4 & 0.958 & 0.077 & 0.041 & 0.006 & 0.173 & 0.039 \\
2000 & 72 & 4 & 4 & 0.965 & 0.033 & 0.039 & 0.004 & 0.140 & 0.025 \\
\bottomrule
\end{tabular}
}
\label{tab:simulation_results_extended}
\end{table}

{We further examined selection of the latent dimensions under $n=1000$ and $J=72$. For each true value $K=2,3,4$, candidate models with $K_1,K_2\in\{1,2,3,4,5\}$ were fitted to 100 simulated datasets. The modified BIC selected the true dimensions $(K_1,K_2)=(K,K)$ in all 100 replicates for each value of $K$. 
Thus, the modified BIC accurately recovered the dimensions of both latent attribute blocks under
the simulation settings considered.
}

\subsection{Comparison with Conventional CDMs}

We further compared the proposed EACDM with a conventional CDM that treats all items as measuring a single latent attribute vector and does not model the directed dependence between the two latent attribute blocks. This comparison examines whether ignoring the endogeneity between $\balpha^{(1)}$ and $\balpha^{(2)}$ distorts recovery of the true latent structure.

We generated 100 datasets with $n=1000$, $J_1=J_2=24$, and $K_1=K_2=3$. The true block-specific $Q$-matrices followed the construction described above, and no observed covariates were included in this comparison. For the structural model, the 
first two endogenous attributes were set to be strongly associated with their corresponding exogenous attributes: the probability of $\alpha_{ik}^{(1)}=1$ was approximately 0.10 when the corresponding component of $\balpha_i^{(2)}$ was 0 and approximately 0.90 when it was 1. The third pair had a weaker dependence, with corresponding probabilities approximately 0.40 and 0.60. This setting creates two highly shared cross-block attribute pairs and one weakly associated pair. Details of the setting including the true parameter values are provided in Supplementary Material.

For the proposed EACDM, we fitted candidate models with $K_1,K_2\in\{2,3,4\}$ and selected the latent dimensions by modified BIC. For the conventional CDM, candidate models with $K\in\{2,3,4,5,6\}$ were fitted, where $K$ denotes the total number of latent attributes in a single unstructured attribute vector. {Under EACDM, the true latent dimensions $(K_1,K_2)=(3,3)$ were correctly selected in all 100 simulation replicates.} In contrast, under the conventional CDM, we rarely selected the true total dimension $K=6$; it selected $K=4$ in 65 replicates, $K=5$ in 33, and $K=6$ in only 2. {The proposed EACDM accurately recovered the block-specific $Q$-matrices and selected the true latent dimensions in all replicates.} In contrast, the conventional CDM typically selected fewer than six attributes, a pattern consistent with the merging of strongly dependent cross-block attributes.

\section{Real-Data Application to the PPMI dataset}
\label{sec:real data}
\subsection{Study Design}

The Parkinson's Progression Markers Initiative (PPMI) is a large longitudinal cohort study designed to characterize Parkinson's disease (PD) progression and identify biomarkers relevant to diagnosis and treatment. We use item-level PPMI assessments to study motor and non-motor manifestations of PD and their cross-domain dependence. The motor domain was constructed from MDS--UPDRS Parts II and III, whereas the non-motor domain was constructed from MDS--UPDRS Part I, SCOPA--AUT, and MoCA items. This partition follows the clinical design of the instruments: MDS--UPDRS Parts II and III primarily measure motor experiences and motor examination findings, while MDS--UPDRS Part I, SCOPA--AUT, and MoCA capture non-motor symptoms, autonomic dysfunction, and cognition.

Motivated by previous evidence linking the caudate and putamen to PD, particularly to its motor manifestations~\citep{RodriguezOroz2009,Pitcher2012}, we include age, caudate uptake, and putamen uptake as covariates in the structural component. The final analytic sample is restricted to participants with complete data on the selected MDS--UPDRS, SCOPA--AUT, and MoCA items and available neuroimaging measurements for the selected brain regions, yielding $n=1127$ participants. Details of data preprocessing are provided in the Supplementary Material.

To improve numerical stability, we excluded items with essentially no response variation. For the remaining items, the most severe response category was rare in several cases; therefore, we collapsed the most severe category with the adjacent less severe category to reduce sparsity in the ordinal responses. The MDS--UPDRS and SCOPA--AUT items are coded so that larger values indicate greater symptom severity, whereas the MoCA items are coded as binary indicators with 1 denoting a correct response and 0 denoting an incorrect response. After preprocessing, the motor domain contains 22 items from MDS--UPDRS Parts II and III. The non-motor domain contains 31 items from MDS--UPDRS Part I, SCOPA--AUT, and MoCA, including sleep, pain, urinary, gastrointestinal, autonomic, sensory, and cognitive items. Item information is provided in Tables~S2 and~S3 of the Supplementary Material.

\subsection{Fitting Results}

\begin{figure}[!htbp]
    \centering
    \includegraphics[
        width=\linewidth,
        height=0.72\textheight,
        keepaspectratio
    ]{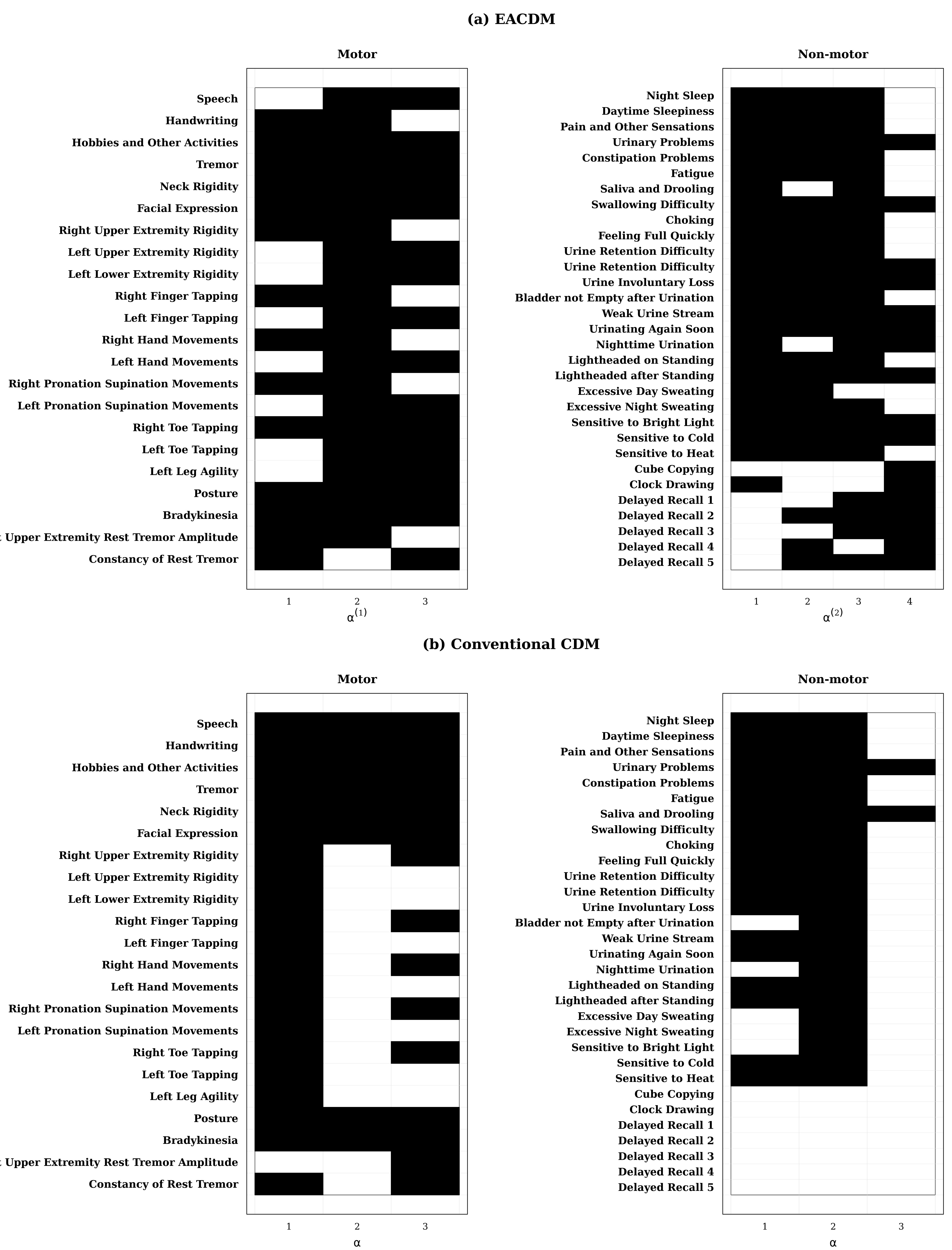}
    \caption{Comparison of the estimated \(Q\)-matrices for the PPMI
    data. Panel (a) shows the block-structured \(Q\)-matrices from the
    EACDM working model with \(K_1=3\) motor attributes and \(K_2=4\)
    non-motor attributes. Panel (b) shows the motor and non-motor
    portions of the \(Q\)-matrix from the conventional CDM with
    \(K=3\). Rows correspond to observed assessment items, columns
    correspond to latent attributes, and black cells indicate selected
    item--attribute relationships based on posterior inclusion
    probabilities thresholded at 0.5.}
    \label{fig:realdata_Q}
\end{figure}

We fitted the proposed EACDM and selected the number of latent attributes using modified BIC together with parsimony and interpretability of the recovered $Q$-matrices. We evaluated the model with different combinations of $(K_1,K_2)$ using BIC and found that BIC tended to decrease as $K_2$ increased, favoring models with more non-motor attributes. However, larger $K_2$ values produced many non-motor latent profiles with low posterior prevalence, resulting in sparsely populated classes that were less reliable for estimating structural relationships between non-motor and motor attributes. We therefore selected a parsimonious and interpretable specification with
$K_1=3$ motor and $K_2=4$ non-motor attributes. Under this specification, all 8 motor and 16 non-motor
marginal profiles had an estimated posterior prevalence of at least 1\%. Panel (a) of Figure~\ref{fig:realdata_Q}
displays the estimated block-specific \(Q\)-matrices.

The estimated motor \(Q\)-matrix has a clinically interpretable
structure. The first motor attribute loads strongly on right-side motor
items, including right upper-extremity rigidity, right finger tapping,
right hand movement, right pronation--supination, right toe tapping,
and right upper-extremity rest tremor, and is therefore interpreted as
right-side motor impairment. The second motor attribute loads broadly
on nearly all motor items, including axial and global motor items such
as speech, facial expression, posture, and bradykinesia, and is
interpreted as global motor impairment. The third motor attribute loads
strongly on left-side motor items, including left upper- and
lower-extremity rigidity, left finger tapping, left hand movement, left
pronation--supination, left toe tapping, and left leg agility, and is
interpreted as left-side motor impairment.

The estimated non-motor \(Q\)-matrix is less clearly separated for its
first three attributes, but it nevertheless provides a useful clinical
structure. These attributes load on broad sets of sleep, pain, urinary,
gastrointestinal, autonomic, and sensory symptoms, suggesting that they
capture overlapping aspects of non-motor and autonomic burden. In
contrast, the fourth non-motor attribute has a clearer interpretation:
it loads strongly on the MoCA items, including cube copying, clock
drawing, and delayed recall. Because the MoCA items are coded as 1 for
correct responses and 0 for incorrect responses, positive measurement
coefficients for these items indicate better cognitive performance. We
therefore interpret \(\alpha^{(2)}_4\) as a cognitive proficiency
attribute. Thus, the non-motor block comprises three broad
non-motor/autonomic burden dimensions and one cognitive proficiency
dimension.

{We next examined the structural coefficients $\boldsymbol{\eta}$, which relate the non-motor latent attributes and observed covariates to the motor latent attributes. Each of the three non-motor/autonomic burden attributes was positively associated with global motor impairment, providing evidence of cross-domain dependence between non-motor burden and motor dysfunction. The third non-motor/autonomic attribute additionally exhibited side-specific
associations: it was negatively associated with right-side motor impairment and positively associated with left-side motor impairment.
By contrast, cognitive proficiency showed no credible association with any motor attribute after adjustment for the other predictors. Among the imaging covariates, higher putamen uptake was associated with lower conditional odds of all three motor impairment attributes, a
pattern consistent with preserved putaminal dopaminergic function. Higher caudate uptake was associated with higher conditional odds of
right-side motor impairment only, whereas age showed no credible association with any motor attribute. Full posterior summaries are provided in Supplementary Table~S4.}

We further examined whether the estimated \(Q\)-matrices were compatible
with the combinatorial requirement in Condition~\((A5')\). For the
motor block, two row-disjoint \(3\times3\) submatrices with unit
diagonals can be selected, while the remaining rows contain at least one
item requiring each of the three attributes. Similarly, for the
non-motor block, two row-disjoint \(4\times4\) submatrices with unit
diagonals can be selected, and each of the four attributes is required
by at least one of the remaining items. Thus, the estimated
\(Q\)-matrix patterns are compatible with the \(Q\)-matrix component of
Condition~\((A5')\), supporting the interpretation of the recovered
latent attributes and their use in the downstream structural analysis.

To further assess the clinical interpretation of the estimated latent
attributes, we used external variables that were not used as primary
item responses for the corresponding validation targets. First, we used
physician-reported dominant side at diagnosis (DOMSIDE) to validate the
side-specific motor attributes. Among subjects classified with the
right-side motor attribute
\((\alpha^{(1)}_1=1)\), 84.3\% had right-dominant onset, compared with
34.6\% among subjects without this attribute. Conversely, among
subjects classified with the left-side motor attribute
\((\alpha^{(1)}_3=1)\), 66.3\% had left-dominant onset, compared with
9.4\% among subjects without this attribute. These patterns support the
interpretation of the first and third motor attributes as right- and
left-side motor impairment, respectively.

We also examined whether the total number of endorsed latent attributes
reflected broader clinical burden. For the motor domain, we used Hoehn
and Yahr stage (NHY) as an external measure of disease severity. The
mean NHY stage increased with motor latent burden, defined as the number
of endorsed motor attributes. The corresponding means were 0.62, 1.46,
1.86, and 2.00 among subjects with 0, 1, 2, and 3 endorsed motor
attributes, respectively.

\begin{figure}[!htbp]
    \centering
    \includegraphics[
        width=0.8\linewidth
    ]{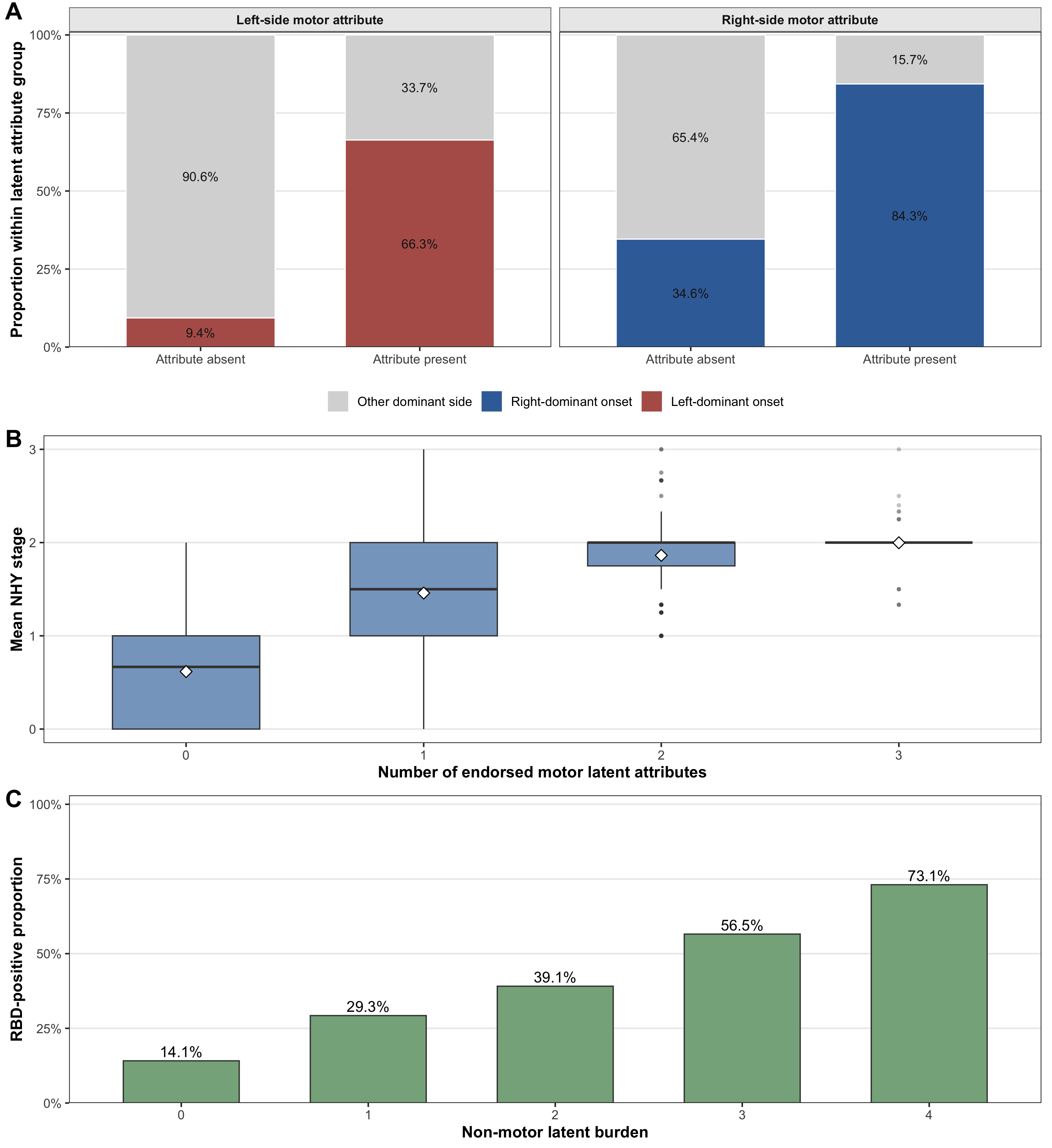}
    \caption{External validation of the estimated latent attributes in
    the PPMI application. Panel A compares physician-reported dominant
    side at diagnosis across the side-specific motor attribute groups.
    Panel B summarizes Hoehn--Yahr stage by the number of endorsed motor
    attributes. Panel C reports the proportion of RBD-positive subjects
    by non-motor latent burden, where the cognitive proficiency
    attribute contributes as \(1-\alpha^{(2)}_4\).}
    \label{fig:external_validation}
\end{figure}

{For the non-motor domain, we used the REM Sleep Behavior Disorder (RBD)
questionnaire as an external validation measure. Because
\(\alpha^{(2)}_4\) represents cognitive proficiency rather than
impairment, non-motor latent burden was defined as
\(
\alpha^{(2)}_1+\alpha^{(2)}_2+\alpha^{(2)}_3+
\{1-\alpha^{(2)}_4\}.
\)
The proportion of RBD-positive subjects increased monotonically with
this non-motor burden, from 14.1\% among subjects with burden 0 to
29.3\%, 39.1\%, 56.5\%, and 73.1\% among those with burdens 1, 2, 3,
and 4, respectively. Taken together, these validation results indicate that
the estimated latent attributes capture clinically meaningful variation in
side-specific motor impairment, overall motor disease severity, and
non-motor symptom burden. Figure~\ref{fig:external_validation}
summarizes the external-validation results.}

\subsection{Comparison with Conventional CDMs}

We further compared the proposed EACDM with a conventional CDM fitted to the same dataset. Unlike the proposed EACDM, the conventional CDM treats all items as measuring a single shared set of latent attributes and does not impose a block structure between the motor and non-motor domains. Panel (b) of Figure~\ref{fig:realdata_Q} shows the estimated $Q$-matrix from the conventional CDM with $K=3$. Several limitations are apparent. First, the estimated attributes do not separate the motor and non-motor domains cleanly. In particular, the first two attributes are associated with items from both domains, suggesting that the conventional CDM combines domain-specific variation into broad shared latent dimensions. Second, the cognitive component is not recovered as a distinct latent attribute. None of the seven MoCA items has a posterior inclusion probability exceeding the 0.5 threshold for any of the three attributes, leaving the cognitive items unassigned in the thresholded $Q$-matrix. Thus, although the conventional CDM captures general response heterogeneity, its three shared attributes provide a less clinically coherent representation of the motor, non-motor, and cognitive structure than the block-specific attributes recovered by EACDM.

\section{Discussion}
\label{sec:discussion}

{Several limitations motivate further work. First, the block-diagonal $Q$-matrix excludes cross-domain item loadings; allowing sparse cross-domain loadings would require regularization and identifiability conditions that distinguish measurement overlap from structural dependence. Second, the present cross-sectional formulation does not separate contemporaneous association from temporal ordering. A longitudinal EACDM could introduce time-varying attributes, transition dynamics, and subject-specific trajectories, while explicitly addressing irregular assessment times and informative dropout. Third, the structural component assumes conditionally independent endogenous attributes given the exogenous block and covariates; residual dependence or reciprocal relationships would require a richer structural model and additional identifying assumptions. Fourth, exact spike-and-slab loading selection and latent-dimension selection introduce uncertainty that is not fully represented by thresholding posterior inclusion probabilities or by the information criterion. Model-averaged inference and sensitivity analyses for slab and selection-prior choices are natural extensions. Finally, MCMC becomes costly as the numbers of items, subjects, or attributes increase; scalable approximations such as variational Bayes warrant investigation.}

\bibliographystyle{plainnat}

\bibliography{references}

\vspace{\fill}

\vfill\eject
\end{document}


\renewcommand{\thetable}{S\arabic{table}}
\renewcommand{\thefigure}{S\arabic{figure}}

\maketitle

\section{Main proofs of identifiability}

{{Recall that the measurement likelihood is exactly $Q$-restricted:} $\mu_{jc}=\beta_{j0}+\sum_kq_{jk}\beta_{jk}\alpha_{ck}=\beta_{j0}+\boldsymbol\delta_j^\top\boldsymbol\alpha_c$, where $\delta_{jk}=q_{jk}\beta_{jk}$ and every active $\beta_{jk}$ is positive. Accordingly, the proofs identify $(\boldsymbol\beta_0,\bcq,\boldsymbol\Delta)$, the exogenous-profile probabilities, and the structural coefficients. All conclusions are understood up to simultaneous permutations of attribute labels within the two blocks.}

\subsection{Proof of Proposition 1}
\begin{proof}
{{For $\mathbf a\in\{0,1\}^{K}$, define $\rho_i(\mathbf a)=P(\boldsymbol\alpha_i=\mathbf a\mid\bcz_i)$. Under the blockwise Condition~$(A5)$, the restricted-latent-class results invoked in the main text identify the category-response vectors $\boldsymbol\theta_{jc}$ and the complete-profile probabilities $\rho_i(\mathbf a)$, up to the stated blockwise label permutations.} \citep{xu2017identifiability,xu2018identifying,culpepper2019exploratory} Condition $(A4)$ and the finite logistic coefficients in $(A2)$ ensure that every complete profile has positive probability, as required for this saturated-profile argument. The dimensional inequality in $(A1)$ is only a necessary check and is not used as a stand-alone sufficient argument.}

{For any item $j$, class $c$, and $m\le M_j-2$, let $F_{jc}(m)=\sum_{r=0}^{m}\theta_{jcr}$. The fixed-threshold probit model gives}
\[
{F_{jc}(m)=\Phi\{\tau_{jc,m+1}-\mu_{jc}\},\qquad
\mu_{jc}=\tau_{jc,m+1}-\Phi^{-1}\{F_{jc}(m)\}.}
\]
{Thus, the identified category probabilities and the injectivity of $\Phi$ identify every class-specific linear predictor $\mu_{jc}$. Condition $(A4)$ supplies the zero and basis profiles of the exogenous block; together with finite logistic coefficients in $(A2)$, these profiles give positive probability to the complete zero profile and to each endogenous and exogenous basis profile needed below. {Under the additive main-effects model, the linear predictor for the all-zero profile equals $\beta_{j0}$, whereas the predictor for the profile with only attribute $k$ present equals $\beta_{j0}+\delta_{jk}$. Thus, the all-zero profile identifies $\beta_{j0}$ and the corresponding single-attribute contrast identifies $\delta_{jk}$. Hence $\boldsymbol\beta_0$ and $\boldsymbol\Delta$ are identified, and positivity gives $q_{jk}=\mathbbm 1\{\delta_{jk}>0\}$ and $\beta_{jk}=\delta_{jk}$ for every active entry.} This identifies $\bcq$ and all active loadings, while no statement is made about an inactive auxiliary $\beta_{jk}$.}

{It remains to identify the structural component. Summing the identified complete-profile probabilities over $\boldsymbol\alpha^{(1)}$ identifies $\pi_{\mathbf a}^{(2)}=P(\boldsymbol\alpha^{(2)}=\mathbf a)$ for every supported exogenous profile $\mathbf a$. Ratios of the identified joint and marginal probabilities then identify $P(\alpha_{ik}^{(1)}=1\mid\boldsymbol\alpha_i^{(2)}=\mathbf a,{\bcz_i})$ for each $k$. Define}
\[
{h_{ik}(\mathbf a)=\operatorname{logit}P(\alpha_{ik}^{(1)}=1\mid\boldsymbol\alpha_i^{(2)}=\mathbf a, {\bcz_i})
=\boldsymbol\lambda_k^\top{\bcz_i}+\boldsymbol\gamma_k^\top\mathbf a.}
\]
{At $\mathbf a=\mathbf0$, Condition $(A3)$ implies that the collection $\{h_{ik}(\mathbf0):i=1,\ldots,N\}$ uniquely determines $\boldsymbol\lambda_k$. For every basis vector $\mathbf e_\ell\in\mathcal S$, $h_{ik}(\mathbf e_\ell)-h_{ik}(\mathbf0)=\gamma_{k\ell}$, so Condition $(A4)$ identifies each component of $\boldsymbol\gamma_k$.} Therefore $(\boldsymbol\beta_0,\bcq,\boldsymbol\Delta,\bm\pi^{(2)},\boldsymbol\gamma,\boldsymbol\lambda)$ is strictly identifiable up to the stated blockwise label permutations.
\end{proof}

\subsection{Proof of Proposition 2}
\begin{proof}
{{Under the blockwise Condition~$(A5')$, the generic-identifiability result for an unknown $Q$-matrix identifies $\bcq$, the category-response vectors $\boldsymbol\theta_{jc}$, and the complete-profile probabilities $\rho_i(\mathbf a)$ outside its exceptional parameter set} \citep{gu2018sufficient}. On that set, the fixed-threshold inversion and zero-and-basis contrasts identify $\boldsymbol\beta_0$ and $\boldsymbol\Delta$; positivity recovers the same active-loading support encoded by $\bcq$. The same marginalization, full-rank, and full-support arguments identify $\bm\pi^{(2)}$, $\boldsymbol\gamma$, and $\boldsymbol\lambda$. The resulting parameterization is therefore generically identifiable up to the stated blockwise label permutations.}
\end{proof}

\section{Simulation settings}

This section provides the structural coefficient values used in the main simulation study. For each simulated subject \(i\), the exogenous latent attribute block \(\boldsymbol{\alpha}_i^{(2)}\) was sampled uniformly from \(\{0,1\}^K\). An independent binary covariate was then generated as
$Z_i\sim \operatorname{Bernoulli}(0.5)$, independently of \(\boldsymbol{\alpha}_i^{(2)}\). Conditional on \(\boldsymbol{\alpha}_i^{(2)}\) and \(Z_i\), the endogenous latent attributes in \(\boldsymbol{\alpha}_i^{(1)}\) were generated independently according to the logistic structural model.

For \(K=2\), the structural coefficient matrix was
\[
\boldsymbol{\eta}_{K=2}
=
\begin{pmatrix}
 0.6 & -0.7\\
 0.8 &  0.8\\
 1.0 & -1.6\\
-0.8 &  0.7
\end{pmatrix}.
\]

For \(K=3\), the structural coefficient matrix was
\[
\boldsymbol{\eta}_{K=3}
=
\begin{pmatrix}
 0.6 & -0.7 & -0.6\\
 0.8 &  0.8 &  0.8\\
 1.0 & -1.6 &  1.4\\
-0.8 &  0.7 & -0.8\\
-1.5 &  1.5 &  0.6
\end{pmatrix}.
\]

For \(K=4\), the structural coefficient matrix was
\[
\boldsymbol{\eta}_{K=4}
=
\begin{pmatrix}
 0.6 & -0.7 & -0.6 &  0.4\\
 0.8 &  0.8 &  0.8 &  0.8\\
 1.0 & -1.6 &  1.4 & -0.6\\
-0.8 &  0.7 & -0.8 &  1.1\\
-1.5 &  1.5 &  0.6 & -0.9\\
 0.7 & -1.0 &  0.8 &  0.9
\end{pmatrix}.
\]
These choices induce a range of positive and negative cross-block dependencies while keeping the covariate effect positive across all endogenous attributes.

\section{Convergence Diagnostics}

\subsection{Convergence diagnostics of simulation studies}

We assessed the convergence of the proposed MCMC algorithm under all
18 simulation settings formed by
\(n\in\{500,1000,2000\}\),
\(J\in\{48,72\}\), and
\(K_1=K_2\in\{2,3,4\}\).
For each setting, we selected replicates 1, 50, and 100 and ran four
independent chains for 3,000 iterations. The first 2,000 iterations were
discarded as burn-in.

Because the latent-attribute labels are exchangeable across chains, the
posterior draws were aligned before computing the convergence
diagnostics. Specifically, for each block, the attribute labels in each
chain were matched to the corresponding true \(Q\)-matrix by minimizing
the Hamming distance. The resulting permutations were applied jointly
to the associated entries of
\(\boldsymbol{\Delta}_1\),
\(\boldsymbol{\Delta}_2\),
\(\boldsymbol{\eta}\),
\(\bcq_1\), and
\(\bcq_2\).
Running Gelman--Rubin statistics were then calculated at checkpoints
from iteration 100 to iteration 3,000. For each scalar parameter, the
running statistics were averaged across the three selected replicates
within each simulation setting.

\begin{sidewaysfigure}[p]
\centering
\includegraphics[width=\textheight]
{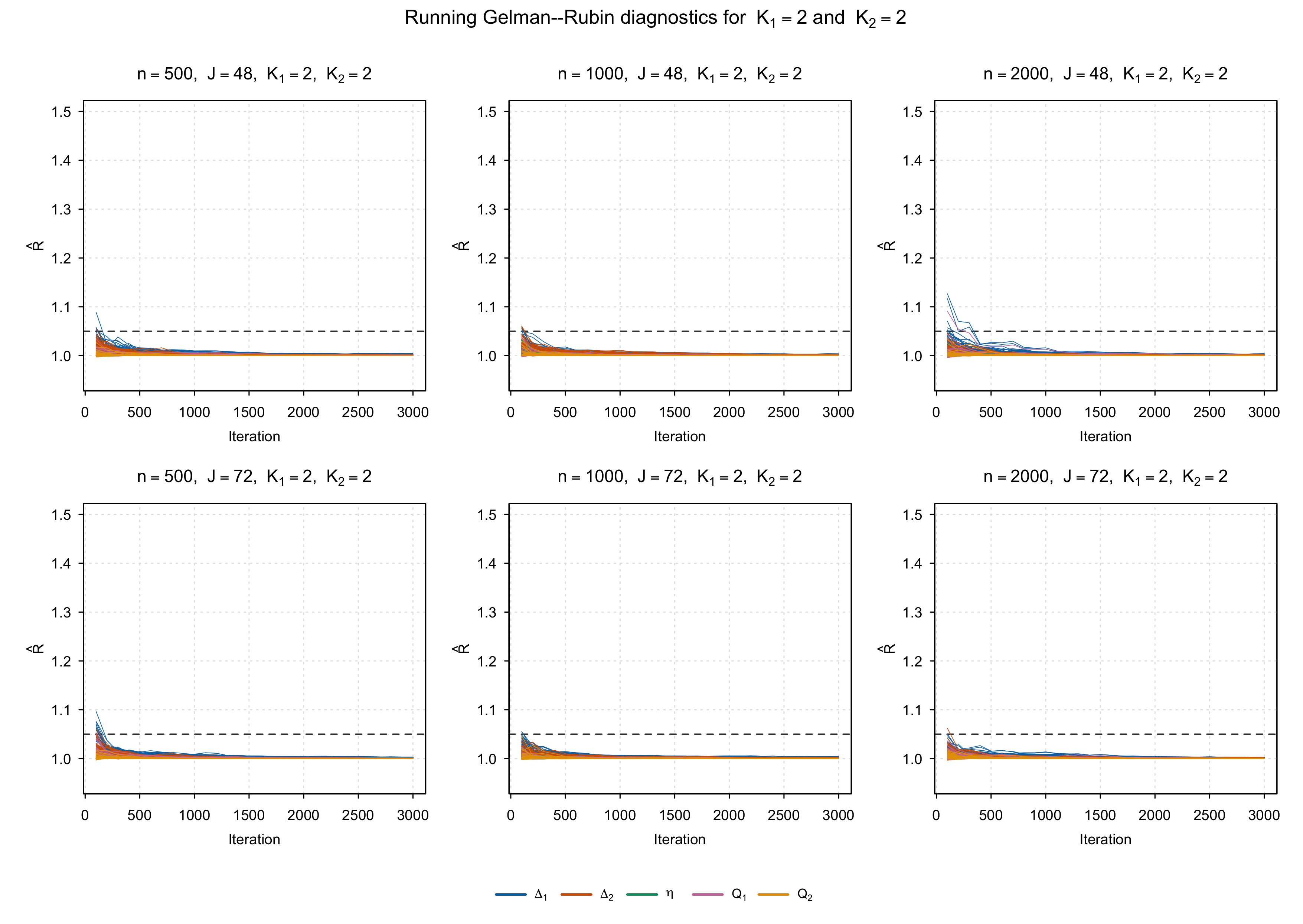}
\caption{Running Gelman--Rubin diagnostics for the simulation settings
with \(K_1=K_2=2\). The curves correspond to
\(\boldsymbol{\Delta}_1\),
\(\boldsymbol{\Delta}_2\),
\(\boldsymbol{\eta}\),
\(\bcq_1\), and
\(\bcq_2\), averaged across the three selected replicates for each
combination of \(n\) and \(J\). The dashed horizontal line denotes
\(\widehat R=1.05\).}
\label{fig:simulation_running_rhat_K2}
\end{sidewaysfigure}

\begin{sidewaysfigure}[p]
\centering
\includegraphics[width=\textheight]
{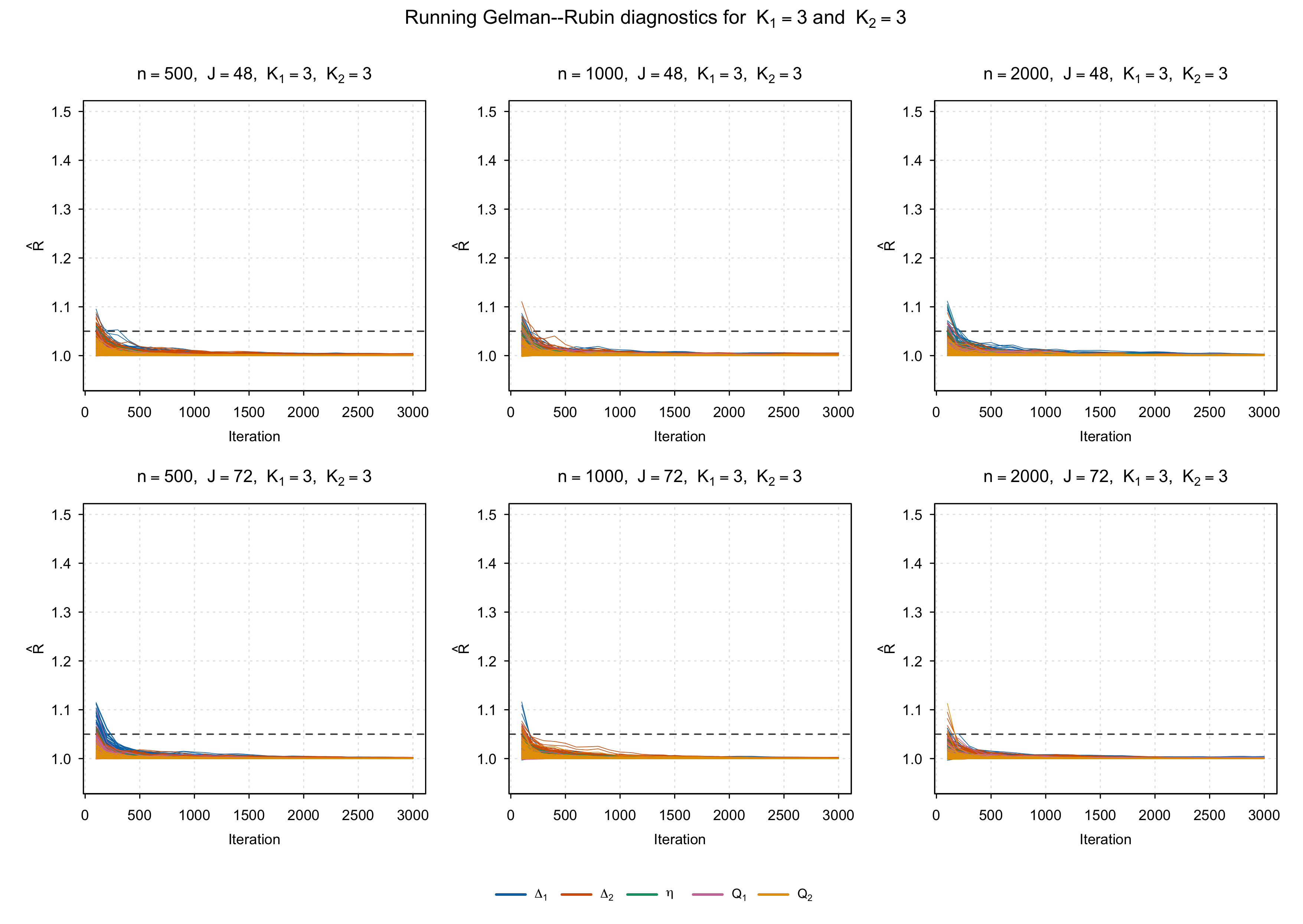}
\caption{Running Gelman--Rubin diagnostics for the simulation settings
with \(K_1=K_2=3\). The curves correspond to
\(\boldsymbol{\Delta}_1\),
\(\boldsymbol{\Delta}_2\),
\(\boldsymbol{\eta}\),
\(\bcq_1\), and
\(\bcq_2\), averaged across the three selected replicates for each
combination of \(n\) and \(J\). The dashed horizontal line denotes
\(\widehat R=1.05\).}
\label{fig:simulation_running_rhat_K3}
\end{sidewaysfigure}

\begin{sidewaysfigure}[p]
\centering
\includegraphics[width=\textheight]
{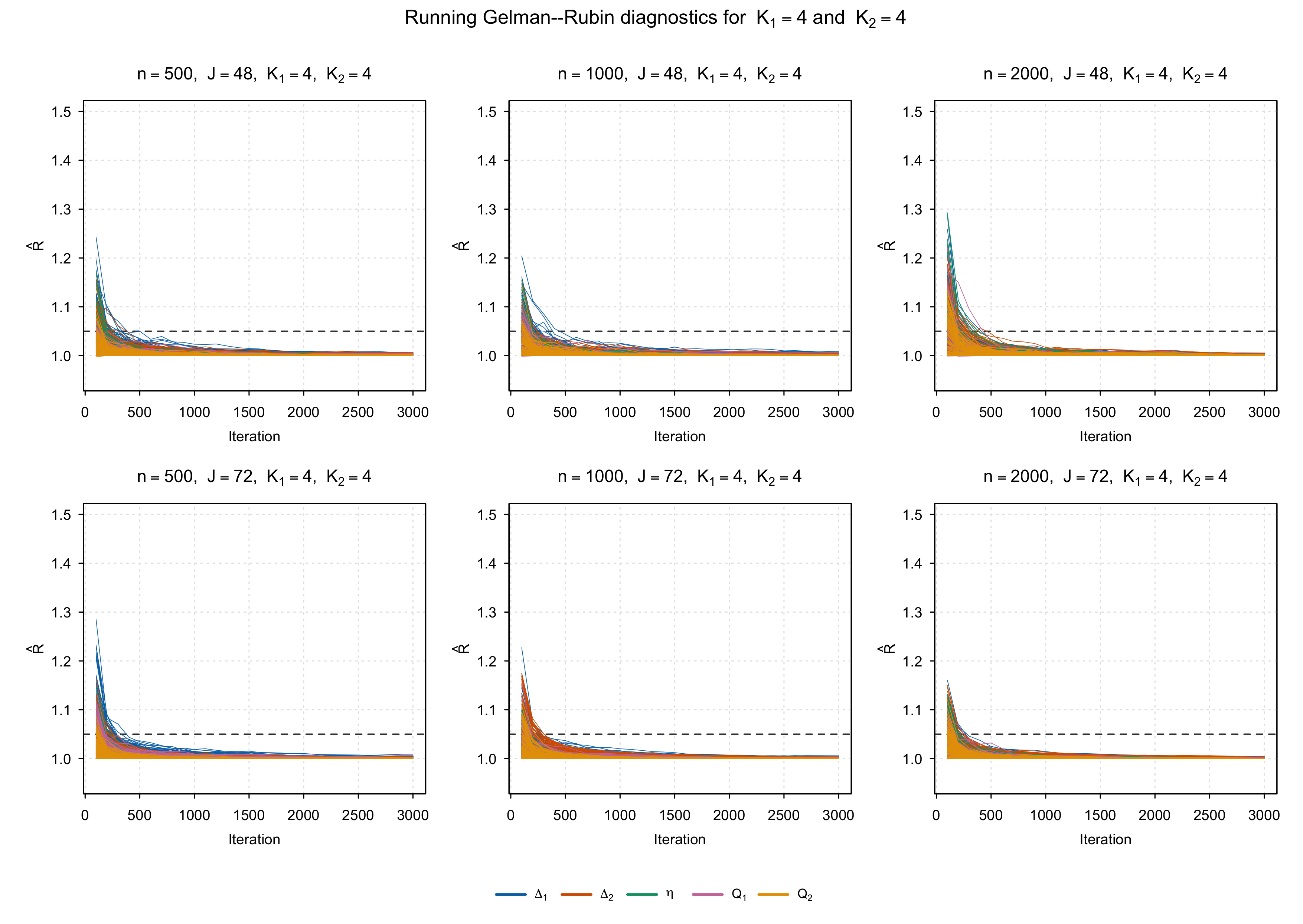}
\caption{Running Gelman--Rubin diagnostics for the simulation settings
with \(K_1=K_2=4\). The curves correspond to
\(\boldsymbol{\Delta}_1\),
\(\boldsymbol{\Delta}_2\),
\(\boldsymbol{\eta}\),
\(\bcq_1\), and
\(\bcq_2\), averaged across the three selected replicates for each
combination of \(n\) and \(J\). The dashed horizontal line denotes
\(\widehat R=1.05\).}
\label{fig:simulation_running_rhat_K4}
\end{sidewaysfigure}

Figures~\ref{fig:simulation_running_rhat_K2}--\ref{fig:simulation_running_rhat_K4} present the resulting running
\(\widehat R\) trajectories for \(K_1=K_2=2,3,\) and \(4\),
respectively. The dashed horizontal line indicates the commonly used
threshold of 1.05.

The running statistics were more variable during the early iterations
but decreased steadily as the chains progressed. The displayed
replicate-averaged trajectories were below 1.05 by iteration 2,000.
At iteration 3,000, the largest displayed average was 1.009. Across all
individual retained-draw diagnostics, 99.89\% of the rank-normalized
split-\(\widehat R\) values were below 1.05. The few larger values were
confined to isolated coordinates of the measurement and \(Q\)-matrix
blocks. Overall, the diagnostics provide evidence of satisfactory
convergence of the proposed sampler across the simulation settings.

\subsection{Convergence diagnostics of the PPMI analysis}

For the PPMI application, we ran four independent chains for 40,000
iterations under the selected model with \(K_1=3\) motor attributes and
\(K_2=4\) non-motor attributes. The first 20,000 iterations were
discarded as burn-in, and iterations 20,001--40,000 were retained for
posterior inference.

As in the simulation study, the latent-attribute labels were aligned
across chains before computing the convergence diagnostics. For each
block, the posterior mean \(Q\)-matrix from the retained draws of the
first chain was used as the reference. The labels in the remaining
chains were matched to this reference by minimizing the Hamming
distance, and the corresponding permutations were applied jointly to
\(\boldsymbol{\Delta}_1\),
\(\boldsymbol{\Delta}_2\),
\(\boldsymbol{\eta}\),
\(\bcq_1\), and
\(\bcq_2\).

\begin{figure}[!htbp]
\centering
\includegraphics[width=\textwidth]
{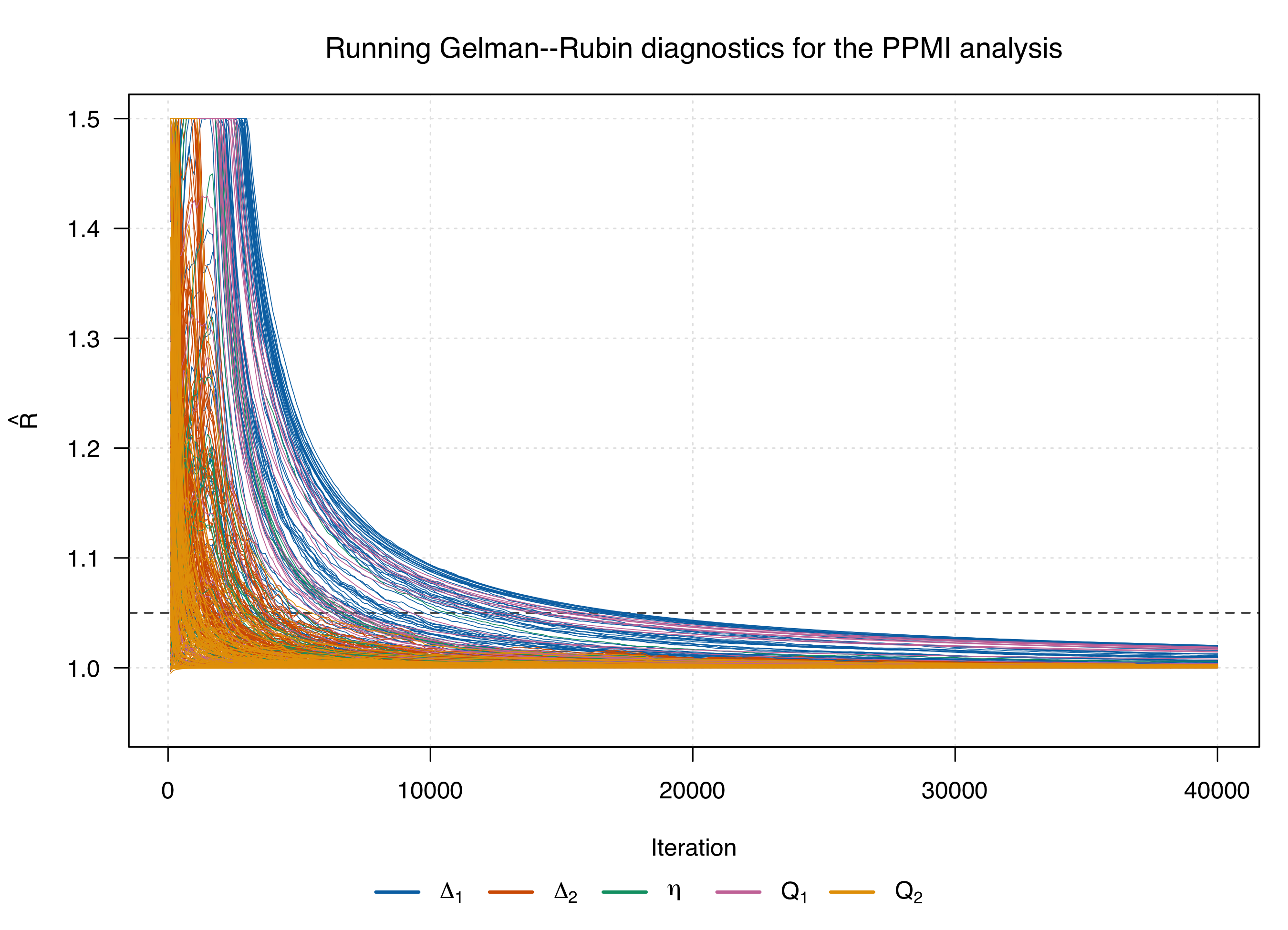}
\caption{Running Gelman--Rubin diagnostics for the EACDM analysis of the
PPMI data under the selected model with \(K_1=3\) and \(K_2=4\).
The curves correspond to
\(\boldsymbol{\Delta}_1\),
\(\boldsymbol{\Delta}_2\),
\(\boldsymbol{\eta}\),
\(\bcq_1\), and
\(\bcq_2\).
The dashed horizontal line denotes \(\widehat R=1.05\).}
\label{fig:ppmi_running_rhat}
\end{figure}

Running Gelman--Rubin statistics were calculated at checkpoints from
iteration 100 to iteration 40,000. Figure~\ref{fig:ppmi_running_rhat}
shows the resulting trajectories. The dashed horizontal line represents
the threshold \(\widehat R=1.05\).

The running \(\widehat R\) trajectories generally decreased toward one
as the number of iterations increased and were below 1.05 by iteration
20,000. Based on the retained iterations 20,001--40,000, the maximum
\(\widehat R\) values were 1.003, 1.007, 1.001, 1.000, and 1.004 for
\(\boldsymbol{\Delta}_1\),
\(\boldsymbol{\Delta}_2\),
\(\boldsymbol{\eta}\),
\(\bcq_1\), and
\(\bcq_2\), respectively. All 457 scalar parameters had
\(\widehat R\leq 1.01\). These results indicate satisfactory convergence
of the posterior sampler for the PPMI analysis.

\section{Simulation Comparison with a Conventional CDM}

We conducted an additional simulation to examine whether a conventional CDM can recover the intended latent structure when two item domains are connected through directed dependence among their latent attributes. We generated 100 independent datasets, each with $n=1000$ subjects, two blocks of $J_1=J_2=24$ items, and two block-specific latent attribute vectors of dimensions $K_1=K_2=3$. All items had three ordered response categories, coded as 0, 1, and 2, and no observed covariates were included.

For each subject $i$, the exogenous attribute vector $\boldsymbol{\alpha}_i^{(2)}\in\{0,1\}^3$ was sampled uniformly from the eight possible binary profiles. Conditional on $\boldsymbol{\alpha}_i^{(2)}$, the three elements of the endogenous attribute vector $\boldsymbol{\alpha}_i^{(1)}$ were generated independently according to
\begin{equation}
\operatorname{logit}\left\{
P\left(\alpha_{ik}^{(1)}=1\mid\boldsymbol{\alpha}_i^{(2)}\right)
\right\}
=\eta_{0k}+\sum_{\ell=1}^{3}\eta_{\ell k}\alpha_{i\ell}^{(2)},
\qquad k=1,2,3,
\label{eq:supp_comparison_structure}
\end{equation}
where the true structural coefficient matrix, with the intercept in the first row, was
\begin{equation}
\boldsymbol{\eta}=
\begin{pmatrix}
-2.20 & -2.20 & -0.40\\
 4.40 &  0    &  0\\
 0    &  4.40 &  0\\
 0    &  0    &  0.80
\end{pmatrix}.
\label{eq:supp_comparison_eta}
\end{equation}
Thus, the dependence was one-to-one: $\alpha_{i1}^{(1)}$, $\alpha_{i2}^{(1)}$, and $\alpha_{i3}^{(1)}$ depended only on $\alpha_{i1}^{(2)}$, $\alpha_{i2}^{(2)}$, and $\alpha_{i3}^{(2)}$, respectively. For the first two pairs, the conditional mastery probabilities were $\operatorname{logit}^{-1}(-2.20)=0.100$ and $\operatorname{logit}^{-1}(2.20)=0.900$ when the corresponding exogenous attribute was 0 and 1. For the third pair, the probabilities were $\operatorname{logit}^{-1}(-0.40)=0.401$ and $\operatorname{logit}^{-1}(0.40)=0.599$. This setting created two strongly dependent cross-block attribute pairs and one weakly dependent pair.

The two item blocks used identical $24\times3$ true $Q$-matrices. Let
\begin{equation}
{\bcq_0}=
\begin{pmatrix}
1&1&0\\
1&0&1\\
0&1&1
\end{pmatrix},
\qquad
{\bcq_e}=
\begin{pmatrix}
{\mathbf I_3}\\
{\mathbf I_3}\\
{\mathbf I_3}\\
{\bcq_0}
\end{pmatrix}.
\label{eq:supp_comparison_qe}
\end{equation}
We set
\begin{equation}
{\bcq_1=\bcq_2}=
\begin{pmatrix}
{\bcq_e}\\
{\bcq_e}
\end{pmatrix}.
\label{eq:supp_comparison_q}
\end{equation}
Consequently, within each 24-item block, each attribute was measured by six single-attribute items, and each of the three pairwise attribute combinations was measured by two items. The complete $48\times6$ loading structure was block diagonal, {$\operatorname{diag}(\bcq_1,\bcq_2)$}.

\subsection{Model fitting and comparison}

For EACDM, we fitted all nine candidate dimension pairs with $K_1,K_2\in\{2,3,4\}$. For the conventional CDM, the two response blocks were concatenated and all 48 items were modeled using one unstructured latent attribute vector. We fitted conventional CDMs with $K\in\{2,3,4,5,6\}$ and included only main attribute effects in the ordinal measurement model. Thus, the conventional CDM was not supplied with the item-block partition and did not include the structural equation in \eqref{eq:supp_comparison_structure}. Each candidate model was estimated using 3,000 MCMC iterations, with the first 2,000 iterations treated as burn-in. For each replicate and method, the candidate with the smallest modified BIC was selected.

{As shown in Table~S1, EACDM selected the true dimensions
\((K_1,K_2)=(3,3)\) in all 100 replicates.} The conventional CDM selected the true total dimension $K=6$ in only 2\% of replicates and most often selected $K=4$.

\begin{table}[ht]
\centering
\caption{BIC-based latent-dimension selection in the simulation comparison.}
\begin{tabular}{llrr}
\toprule
Model & Selected latent dimension & Count & Proportion \\
\midrule
{EACDM} & $(K_1,K_2)=(3,3)$ & 100 & 1.00 \\
\midrule
Conventional CDM & $K=4$ & 65 & 0.65 \\
Conventional CDM & $K=5$ & 33 & 0.33 \\
Conventional CDM & $K=6$ &  2 & 0.02 \\
\bottomrule
\end{tabular}
\label{tab:supp_model_selection}
\end{table}

\begin{figure}[p]
\centering
\includegraphics[width=0.82\linewidth]{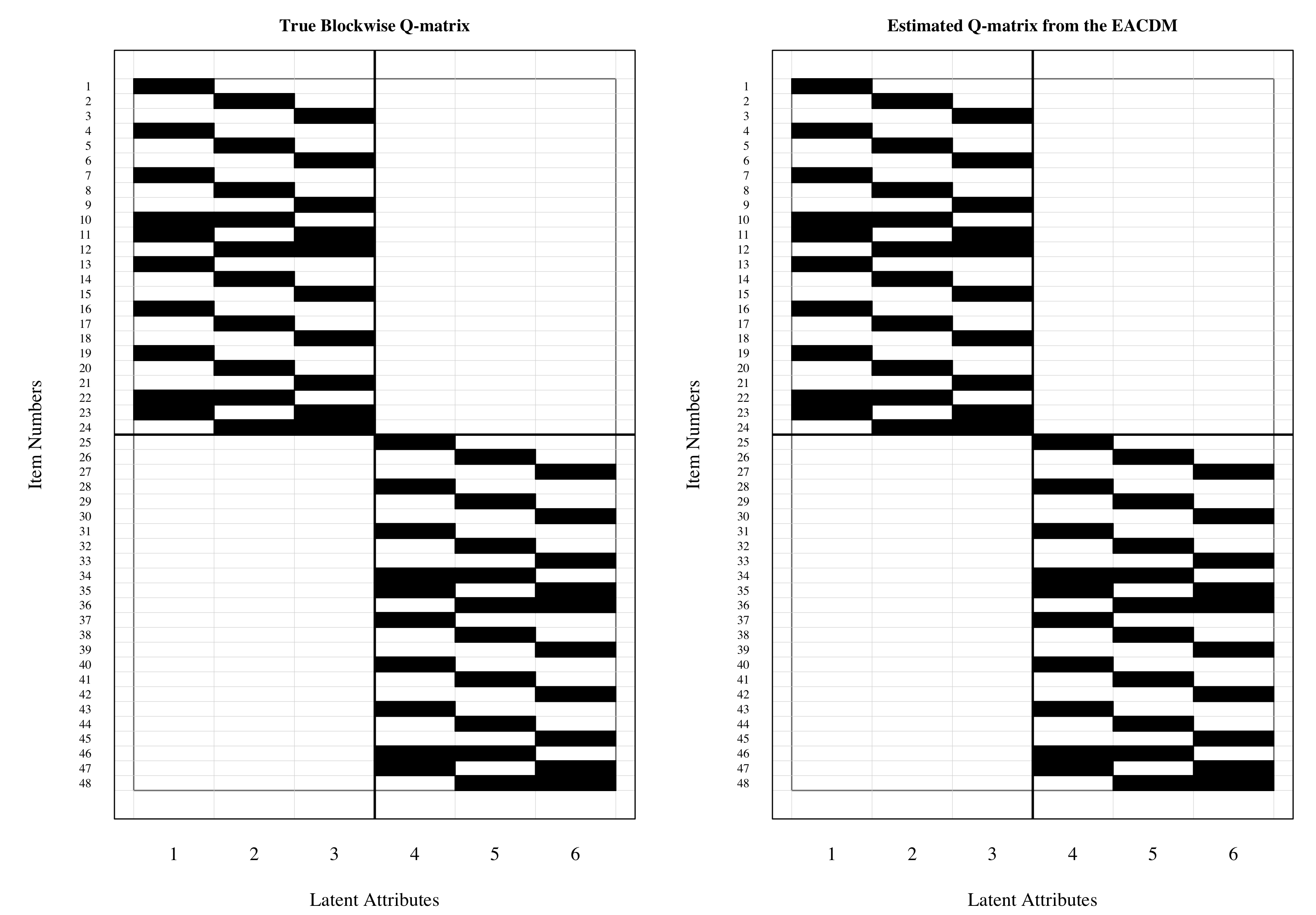}
\caption{Complete true and estimated blockwise $Q$-matrices for {EACDM} in the simulation comparison. The posterior mean matrix is averaged over the replicates with the modal selected dimensions $(K_1,K_2)=(3,3)$ and thresholded at 0.5. The horizontal line separates the two item blocks, and the vertical line separates their block-specific latent attributes. Dark cells indicate active item--attribute relationships.}
\label{fig:simulation_secdm_q}
\end{figure}

\begin{figure}[p]
\centering
\includegraphics[width=0.82\linewidth]{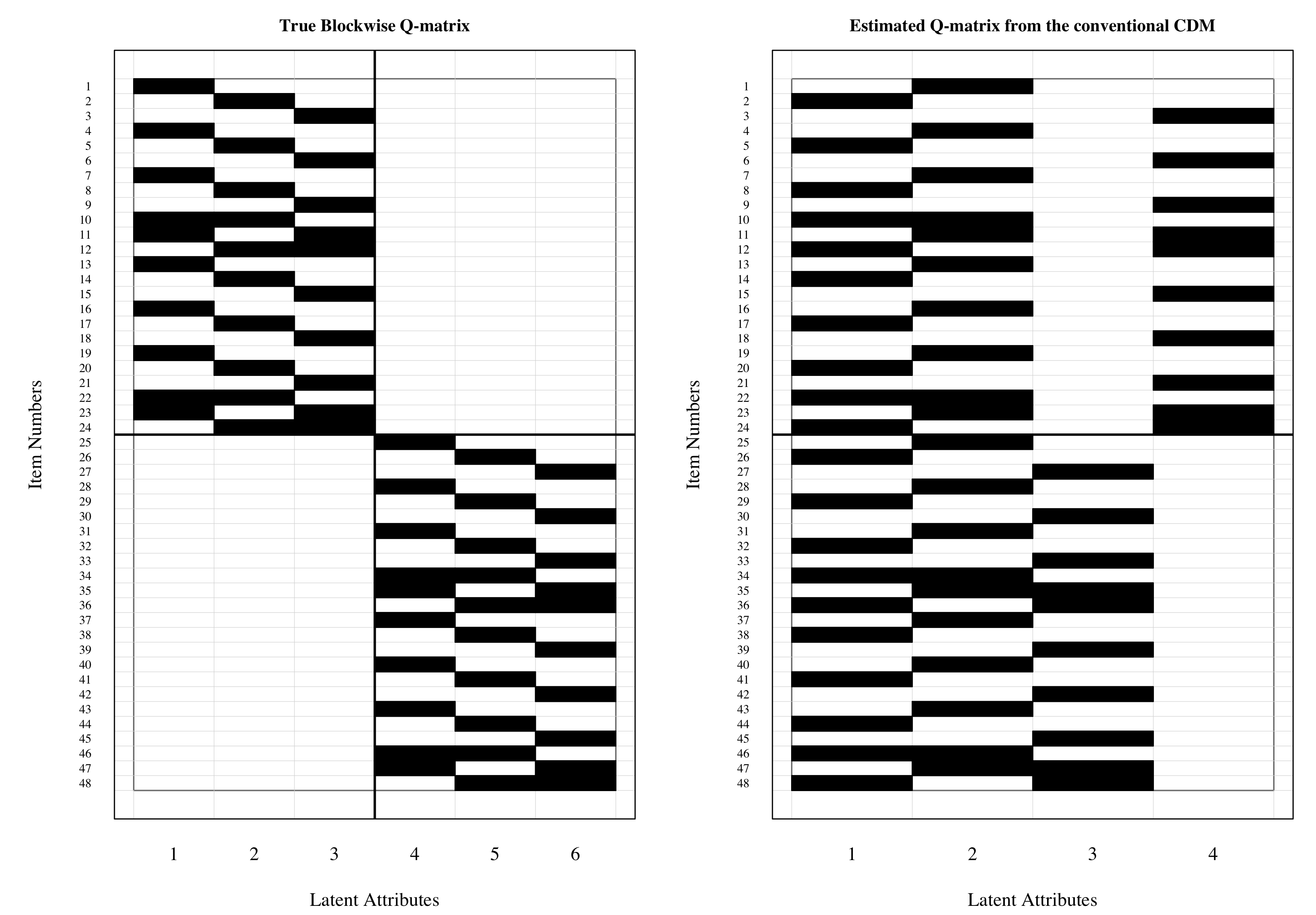}
\caption{Complete true blockwise $Q$-matrix and the estimated $Q$-matrix from the conventional CDM. The posterior mean matrix is averaged over the replicates with the modal selected dimension $K=4$ and thresholded at 0.5. The horizontal line separates the two item blocks. Unlike {EACDM}, the conventional CDM uses four common latent attributes for all items and therefore has no block-specific column partition. Dark cells indicate active item--attribute relationships.}
\label{fig:simulation_cdm_q}
\end{figure}

Figures~\ref{fig:simulation_secdm_q} and~\ref{fig:simulation_cdm_q} compare the complete true blockwise $Q$-matrix with the corresponding posterior mean estimate, averaged over replicates having the modal selected dimensions. The horizontal line separates the two 24-item blocks. In the true and {EACDM} matrices, the vertical line separates the two sets of three block-specific latent attributes. The {EACDM} result closely reproduces this block-diagonal structure. In contrast, the conventional CDM represents all 48 items using four common latent attributes and mixes item--attribute patterns across the two item blocks. The comparison illustrates why an unstructured CDM can underestimate the effective latent dimension when strongly dependent attributes are represented as a single attribute vector.

\section{Data preprocessing in PPMI application}
Because the assessment schedules in PPMI differ across instruments, requiring MDS--UPDRS, SCOPA--AUT, and MoCA to be available at the same visit would be overly restrictive and would lead to a substantial loss of participants. We therefore used a closely spaced early-visit window consisting of Screening (SC), Baseline (BL), Visit 01 (V01), and Visit 02 (V02). According to the PPMI visit schedule, SC and BL are adjacent early study visits, V01 is scheduled approximately 3 months after baseline, and V02 approximately 6 months after baseline. Thus, although the selected assessments were not always administered at the same nominal visit, they were taken from an early period of follow-up in which disease progression is expected to have a more limited impact than over later visits. 

The eligible visits differed by instrument because the three assessments were not administered at exactly the same set of visits. MDS--UPDRS records were selected from SC, V01, or V02, whereas SCOPA--AUT and MoCA records were selected from SC, BL, V01, or V02. After applying the visit and completeness filters, one eligible record per participant was retained for each assessment. In the final analytic sample, MDS--UPDRS records came from SC for 49 participants, V01 for 445 participants, and V02 for 633 participants. SCOPA--AUT records came from BL for 1,124 participants and V02 for 3 participants, while MoCA records came from SC for 1,126 participants and BL for 1 participant. Consequently, the analytic dataset primarily aligned MDS--UPDRS assessments from V01 or V02 with SCOPA--AUT assessments from BL and MoCA assessments from SC. This early-window strategy allowed us to retain participants whose assessments were administered on different PPMI schedules while limiting the temporal separation among the measurements.

\clearpage
\section{Supplementary Tables for the PPMI Application}

The following tables list the item stems and empirical response-category proportions for the motor and non-motor item sets used in the real-data analysis.

\begin{table}[!htbp]
\centering
\scriptsize
\setlength{\tabcolsep}{4pt}
\caption{Motor part: Selected MDS--UPDRS item stems and response-category proportions after preprocessing {in the analytic sample of $n=1{,}127$ participants}.}
\begin{tabular}{lcccc}
\toprule
\textbf{Item} & \textbf{0} & \textbf{1} & \textbf{2} & \textbf{3} \\
\midrule
1. Speech & 0.6389 & 0.2289 & 0.1145 & 0.0177 \\
2. Handwriting & 0.3940 & 0.3469 & 0.1748 & 0.0843 \\
3. Doing hobbies and other activities & 0.5909 & 0.2848 & 0.0923 & 0.0320 \\
4. Tremor & 0.2014 & 0.5359 & 0.2245 & 0.0382 \\
5. Rigidity (neck) & 0.4933 & 0.3052 & 0.1801 & 0.0214 \\
6. Facial expression & 0.2121 & 0.5058 & 0.2564 & 0.0257 \\
7. Rigidity (RUE) & 0.3301 & 0.3416 & 0.2990 & 0.0293 \\
8. Rigidity (LUE) & 0.4374 & 0.3159 & 0.2236 & 0.0231 \\
9. Rigidity (LLE) & 0.6193 & 0.2343 & 0.1304 & 0.0160 \\
10. Finger tapping (R) & 0.3194 & 0.3762 & 0.2351 & 0.0693 \\
11. Finger tapping (L) & 0.3611 & 0.3097 & 0.2449 & 0.0843 \\
12. Hand movements (R) & 0.4286 & 0.3665 & 0.1659 & 0.0390 \\
13. Hand movements (L) & 0.4330 & 0.3372 & 0.1854 & 0.0444 \\
14. Pronation-supination movements of hand (R) & 0.4286 & 0.3656 & 0.1650 & 0.0408 \\
15. Pronation-supination movements of hand (L) & 0.4587 & 0.3088 & 0.1819 & 0.0506 \\
16. Toe tapping (R) & 0.4082 & 0.3860 & 0.1579 & 0.0479 \\
17. Toe tapping (L) & 0.3780 & 0.3452 & 0.2165 & 0.0603 \\
18. Leg agility (L) & 0.5812 & 0.2848 & 0.1198 & 0.0142 \\
19. Posture & 0.4623 & 0.4108 & 0.1162 & 0.0107 \\
20. Body bradykinesia & 0.2085 & 0.4437 & 0.2964 & 0.0514 \\
21. Rest tremor amplitude (RUE) & 0.6238 & 0.1384 & 0.1766 & 0.0612 \\
22. Constancy of rest tremor & 0.3780 & 0.2449 & 0.1562 & 0.2209 \\
\bottomrule
\end{tabular}
\vspace{0.5em}
\noindent\parbox{\textwidth}{\footnotesize Note: The anchor labels before collapsing were 0 = normal, 1 = slight, 2 = mild, 3 = moderate, and 4 = severe; the severe category was collapsed into category 3 for analysis.}
\label{tab:motor_items}
\end{table}

\begin{table}[!htbp]
\centering
\scriptsize
\caption{Non-motor part: Selected MDS--UPDRS, SCOPA, and MoCA item stems and response-category proportions after preprocessing {in the analytic sample of $n=1{,}127$ participants}.}
\begin{tabular}{lcccc}
\toprule
\textbf{Item} & \textbf{0} & \textbf{1} & \textbf{2} & \textbf{3} \\
\midrule
23. Sleep problems & 0.4295 & 0.2626 & 0.1801 & 0.1278 \\
24. Daytime sleepiness & 0.4463 & 0.2937 & 0.2351 & 0.0249 \\
25. Pain and other sensations & 0.4508 & 0.3594 & 0.1189 & 0.0709 \\
26. Urinary problems & 0.5146 & 0.3372 & 0.1012 & 0.0470 \\
27. Constipation problems & 0.6344 & 0.2662 & 0.0728 & 0.0266 \\
28. Fatigue & 0.4685 & 0.3718 & 0.1162 & 0.0435 \\
29. Saliva and drooling  & 0.6477 & 0.1473 & 0.1287 & 0.0763 \\
30. Had difficulty swallowing & 0.7604 & 0.2130 & 0.0266 & / \\
31. Had been choked  & 0.8163 & 0.1704 & 0.0133 & / \\
32. Have the feeling during a meal that you were full very quickly & 0.7471 & 0.2112 & 0.0417 & / \\
33. Have to strain hard to pass stools & 0.4862 & 0.3895 & 0.1243 & / \\
34. Had difficulty retaining urine & 0.6371 & 0.2813 & 0.0816 & / \\
35. Had involuntary loss of urine & 0.7125 & 0.2520 & 0.0355 & / \\
36. Had feeling that after passing urine the bladder was not completely empty & 0.5679 & 0.3319 & 0.1002 & / \\
37. Has the stream of urine been weak & 0.5563 & 0.3310 & 0.1127 & / \\
38. Had to pass urine again within 2 hours of the previous time & 0.1917 & 0.5581 & 0.2502 & / \\
39. Had to pass urine at night & 0.1668 & 0.3638 & 0.4694 & / \\
40. When standing up had the feeling of becoming light-headed & 0.6939 & 0.2751 & 0.0310 & / \\
41. Become light-headed after standing for some time & 0.8554 & 0.1287 & 0.0159 & / \\
42. Perspired excessively during the day & 0.7986 & 0.1464 & 0.0550 & / \\
43. Perspired excessively during the night & 0.7409 & 0.2094 & 0.0497 & / \\
44. Eyes ever been over-sensitive to bright light & 0.6877 & 0.2413 & 0.0710 & / \\
45. Had trouble tolerating cold & 0.6593 & 0.2484 & 0.0923 & / \\
46. Had trouble tolerating heat & 0.7516 & 0.1837 & 0.0647 & / \\
47. Copy cube & 0.2547 & 0.7453 & / & / \\
48. Draw clock & 0.1899 & 0.8101 & / & / \\
49. {Delayed recall 1} & 0.4161 & 0.5839 & / & / \\
50. {Delayed recall 2} & 0.2298 & 0.7702 & / & / \\
51. {Delayed recall 3} & 0.2866 & 0.7134 & / & / \\
52. {Delayed recall 4} & 0.4348 & 0.5652 & / & / \\
53. {Delayed recall 5} & 0.2999 & 0.7001 & / & / \\
\bottomrule
\end{tabular}
\vspace{0.5em}
\noindent\parbox{\textwidth}{\footnotesize Note: Items 23--28 are from MDS--UPDRS and labels before collapsing were 0 = normal, 1 = slight, 2 = mild, 3 = moderate, and 4 = severe; the severe category was collapsed into category 3 for analysis. Items 29--46 are from SCOPA and labels before collapsing were 0 = never, 1 = sometimes, 2 = regularly, and 3 = often; the often category was collapsed into category 2 for analysis. Items 47--53 are from MoCA and labels were 0 = wrong and 1 = correct.}
\label{tab:nonmotor_items}
\end{table}

\clearpage

In addition to the item-level summaries, we report posterior summaries of the structural coefficients $\boldsymbol{\eta}$ from the selected PPMI model. These coefficients correspond to the logistic structural component linking the non-motor latent attributes and covariates to the motor latent attributes. Thus, each coefficient is interpreted on the log-odds scale: positive values indicate higher posterior probability of endorsing the corresponding motor attribute, conditional on the other predictors, whereas negative values indicate lower posterior probability. Table~S4 reports the posterior mean, 95\% credible interval, and posterior probability of a positive effect for each coefficient.

\begin{table}[!htbp]
\centering
\caption{Posterior summaries of the structural coefficients
\(\boldsymbol{\eta}\) in the selected PPMI model with \(K_1=3\)
motor attributes and \(K_2=4\) non-motor attributes. The motor
attributes are ordered as right-side motor impairment, global motor
impairment, and left-side motor impairment. The non-motor attributes
\(\alpha^{(2)}_1\), \(\alpha^{(2)}_2\), and \(\alpha^{(2)}_3\)
correspond to broad non-motor/autonomic burden attributes, whereas
\(\alpha^{(2)}_4\) corresponds to cognitive proficiency. Positive
coefficients indicate increased log-odds of endorsing the corresponding
motor attribute.}
\label{tab:ppmi_eta}
\small
\setlength{\tabcolsep}{7pt}
\begin{tabular}{llrrrr}
\toprule
Predictor & Motor attribute & Mean & 2.5\% & 97.5\% &
\(\Pr(\eta>0)\) \\
\midrule
Intercept
& Right-side motor & -0.638 & -1.426 & 0.151 & 0.056 \\
\(\alpha^{(2)}_1\)
& Right-side motor & 0.184 & -0.162 & 0.535 & 0.847 \\
\(\alpha^{(2)}_2\)
& Right-side motor & -0.239 & -0.645 & 0.163 & 0.120 \\
\(\alpha^{(2)}_3\)
& Right-side motor & -0.788 & -1.228 & -0.361 & 0.000 \\
\(\alpha^{(2)}_4\)
& Right-side motor & -0.335 & -0.721 & 0.050 & 0.044 \\
Age
& Right-side motor & 0.838 & -0.271 & 1.946 & 0.931 \\
Caudate uptake
& Right-side motor & 0.383 & 0.128 & 0.641 & 0.998 \\
Putamen uptake
& Right-side motor & -0.571 & -0.855 & -0.292 & 0.000 \\
\midrule
Intercept
& Global motor & -0.985 & -1.865 & -0.118 & 0.013 \\
\(\alpha^{(2)}_1\)
& Global motor & 0.717 & 0.283 & 1.156 & 0.999 \\
\(\alpha^{(2)}_2\)
& Global motor & 0.557 & 0.102 & 1.011 & 0.992 \\
\(\alpha^{(2)}_3\)
& Global motor & 1.370 & 0.835 & 1.940 & 1.000 \\
\(\alpha^{(2)}_4\)
& Global motor & 0.121 & -0.314 & 0.563 & 0.705 \\
Age
& Global motor & 0.082 & -1.132 & 1.292 & 0.552 \\
Caudate uptake
& Global motor & 0.145 & -0.144 & 0.437 & 0.837 \\
Putamen uptake
& Global motor & -1.216 & -1.595 & -0.854 & 0.000 \\
\midrule
Intercept
& Left-side motor & -0.743 & -1.566 & 0.078 & 0.038 \\
\(\alpha^{(2)}_1\)
& Left-side motor & 0.202 & -0.178 & 0.582 & 0.850 \\
\(\alpha^{(2)}_2\)
& Left-side motor & -0.073 & -0.501 & 0.353 & 0.369 \\
\(\alpha^{(2)}_3\)
& Left-side motor & 1.346 & 0.873 & 1.842 & 1.000 \\
\(\alpha^{(2)}_4\)
& Left-side motor & 0.206 & -0.201 & 0.619 & 0.839 \\
Age
& Left-side motor & 0.639 & -0.517 & 1.797 & 0.860 \\
Caudate uptake
& Left-side motor & 0.094 & -0.174 & 0.363 & 0.752 \\
Putamen uptake
& Left-side motor & -0.842 & -1.149 & -0.546 & 0.000 \\
\bottomrule
\end{tabular}
\end{table}

\label{lastpage}

\clearpage

\bibliographystyle{plainnat}
\bibliography{references}